%% file: stat-sinica2.tex
\documentclass[12pt]{article}
\include{preamble}
\begin{document}
\thispagestyle{empty} \single \bc {\bf \sc \Large Derivative Principal Component Analysis for Representing the Time Dynamics of Longitudinal and Functional Data}\footnote{Research
supported  by
 NSF grant  DMS-1407852.}\\ \vspace{0.25in}

Short title: Derivative Principal Component Analysis \\[2em]

\textit{Accepted manuscript by Statistica Sinica}\\[2em]

Xiongtao Dai,
Hans-Georg M\"uller\\
Department of Statistics\\
University of California\\
Davis, CA 95616 USA\\[.1in]

and\\[.1in]

Wenwen Tao\\
Quora\\
Mountain View, CA 94041 USA \ec

\vspace{.02in} \centerline{\today} \vspace{.3in}

\pagenumbering{arabic} \setcounter{page}{0} 

\vspace{-.25in} \thispagestyle{empty} \bc {\bf \sf ABSTRACT} \ec
\vspace{-0.05in}

\double
\no  We propose a nonparametric method to explicitly model and represent the derivatives of smooth underlying trajectories for longitudinal data. This representation is  based on a direct  Karhunen--Lo\`eve expansion of the unobserved derivatives and leads to the notion of derivative principal component analysis, which complements functional principal component analysis, one of the most popular tools of functional data analysis. The proposed derivative principal component scores  can be obtained for irregularly spaced and sparsely observed longitudinal data, as typically encountered in biomedical studies,  as well as for functional data which are densely measured. Novel consistency results and asymptotic convergence rates for the proposed estimates of the derivative principal component scores and other components of the model are derived under a unified scheme for sparse or dense observations and mild conditions. We compare the proposed representations for derivatives with alternative approaches in simulation settings and also in a wallaby growth curve application. It emerges that representations using the proposed derivative principal component analysis recover the underlying derivatives more accurately compared to principal component analysis-based approaches especially in settings where the functional data are represented with only a very small number of components or are densely sampled.  In a second  wheat spectra classification example, derivative principal component scores were found to be more predictive for the protein content of wheat than the conventional functional principal component scores.\\

\no {Keywords:  Derivatives, Empirical dynamics, Functional principal component analysis, Growth curves, Best linear unbiased prediction.}\thispagestyle{empty} \vfill \vskip0.15in
\noindent

\newpage
\section{\sf Introduction} \label{s:intro}

Estimating derivatives and representing the dynamics for longitudinal data is often crucial for  a better description and understanding of the time dynamics that generate longitudinal data \cp{mull:10:2}. Representing  derivatives, however, is not straightforward. Efficient representations of derivatives can be based on expansions into eigenfunctions of derivative processes. Difficulties abound in scenarios with  sparse designs and noisy data. To address these issues,  we propose a method for representing the derivatives of observed longitudinal data by directly aiming at the Karhunen--Lo\`eve expansion \cp{gren:50}   of derivative processes.  Classical methods for estimating derivatives of random trajectories usually require observed data to be densely sampled. These methods include difference quotients, estimates based on B-splines \cp{debo:72}, smoothing splines \cp{cham:91, zhou:00}, kernel-based estimators such as convolution-type kernel estimators \cp{mull:84:1}, and local polynomial estimators \cp{fan:96}. In the case of sparsely and irregularly  observed data, however, direct estimation of derivatives for each single function is not feasible due to the gaps in the measurement times.

For the case of  irregular and sparse designs,  \ci{mull:09:1} proposed a  method based on functional principal component analysis (FPCA)  \cp{rice:91, rams:05} for estimating derivatives. The central  idea of FPCA is dimension reduction by means of a  spectral decomposition of the autocovariance operator, which yields functional principal component scores (FPCs) as coefficient vectors to represent the random curves in the sample.
In \ci{mull:09:1},  derivatives of eigenfunctions are estimated and  plugged in to obtain derivatives of the  estimated Karhunen--Lo\`eve representation for  the random trajectories.
While this method was shown to outperform several  other approaches for recovering derivatives for sparse and irregular designs,  including those using difference quotients, functional mixed effect models with B-spline basis functions \cp{shi:96,rice:01}, or P-splines \cp{jank:05:1,redd:06, bapn:08,wang:08:2}, it is suboptimal for representing derivatives, as the coefficients in the Karhunen--Lo\`eve expansion are targeting the functions themselves and not the derivatives. 

This provides the key motivation for this paper: represent dynamics by directly targeting the Karhunen--Lo\`eve representation of derivatives  of random trajectories. The Karhunen--Lo\`eve representation of derivatives needs to be inferred from available data, which are modeled as noisy measurements of the trajectories.  We then aim to represent derivatives directly in  their corresponding eigenbasis, yielding the most parsimonious representation, and leading to a novel  Derivative Principal Component Analysis (DPCA). In addition, the resulting expansion coefficients, which we refer to as derivative principal component scores  (DPCs) provide a novel representation and dimension reduction tool for functional data that
complements other such representations such as the commonly used FPCs.

The proposed method is designed for both sparse and dense cases and works successfully under both cases. When the functional data are densely sampled with possibly large measurement errors, smoothing  the observed trajectories and obtaining derivatives  for  each trajectory separately is subject to possibly large estimation errors, which are further amplified for derivatives. In contrast, the proposed method  pools observations across subjects and utilizes information from measurements at nearby time points from all subjects when targeting derivatives, and therefore is less affected by large measurement errors. In  scenarios where only a few measurements are available for each subject, the proposed method  performs derivative estimation by borrowing strength from all observed data points, instead of  relying on the sparse data that are observed for a specific trajectory. A key step  is to model and estimate the eigenfunctions of the random derivative functions directly, by spectral-decomposing the covariance function of the derivative trajectories.

The main novelty of our work is to obtain empirical Karhunen--Lo\`eve representations for the dynamics of both sparsely measured longitudinal data and densely measured functional data, and to obtain the DPCA with corresponding DPCs. For the estimation of these DPCs, we employ a best linear unbiased prediction (BLUP) method that directly predicts the DPCs based on the observed measurements. In the special case of a Gaussian process with independent Gaussian noises, the BLUP method coincides with the best prediction.  This unified approach provides a straightforward implementation for the Karhunen--Lo\`eve representation of derivatives. Under a unified framework for the sparse and the dense case, we provide convergence rate results for the derivatives of the mean function, the covariance function, and the derivative eigenfunctions based on smoothing the pooled scatter plots \citep{zhan:16}. We also derive convergence rates for the estimated DPCs based on BLUP. %, where the proof of the convergence rate results for the DPCs in the dense case overcomes some technical challenges. 

The remainder of this paper is structured as follows: In \autoref{s:rep}, we introduce the new representations for derivatives. DPCs and their estimation are the topic of \autoref{s:est}.  Asymptotic properties of the estimated components and of the resulting derivative estimates are presented in \autoref{s:asymp}. We compare the proposed method with alternative approaches in terms of derivatives recovery in \autoref{s:sim} via simulation studies and in \autoref{s:application} using longitudinal wallaby body length data. As is demonstrated in \autoref{s:application}, DPCs can be used to improve classification of functional data, illustrated by wheat spectral data. Additional details are provided in the Appendix. 

\section{\sf Representing Derivatives of Random Trajectories} \label{s:rep}
\subsection{Preliminaries} \label{ss:prelim}

Consider a $\nu$-times differentiable stochastic process $X$ on a compact interval $\cT \subset \bbR$, with $X^{(\nu)} \in L^2(\cT)$, mean function $E(X(t))=\mu(t)$, and auto-covariance function $\cov(X(s), X(t))=G(s,t)$, for $s,t \in \mathcal{T}$. The independent realizations $X_1, \dots, X_n$ of $X$ can be represented in  the Karhunen--Lo\`eve expansion,
\begin{equation}
\displaystyle X_i(t) = \mu(t) +\sum_{k=1}^{\infty}\xi_{ik}\phi_{k}(t), \label{eqn: X_0 KL}
  \end{equation}
where $ \displaystyle \xi_{ik}=\int\left(X_i(t)-\mu(t)\right)\phi_{k}(t)dt$ are the functional principal component scores (FPCs) of the random functions $X_i$ that satisfy $E(\xi_{ik})=0$, $E(\xi_{ik}^2)=\lambda_k$, $E(\xi_{ik}\xi_{ij})=0$ for $k,j=1, 2, \dots$, $k\ne j$; the $\phi_k$ are the eigenfunctions of the covariance operator associated with $G$, with ordered eigenvalues $\lambda_1>\lambda_2>\ldots\geq0$.   

By taking the $\nu$th derivative with respect to  $t$ on both sides of \eqref{eqn: X_0 KL},  \ci{mull:09:1} 
obtained a representation of derivatives,
\begin{equation}
\displaystyle X_i^{(\nu)}(t) = \mu^{(\nu)}(t) +\sum_{k=1}^{\infty}\xi_{ik}\phi_{k}^{(\nu)}(t), \label{eqn: X_1 KL}
\end{equation}
assuming that both sides are well defined, 
with corresponding variance $\var(X_i^{(\nu)}(t))=\sum_{k=1}^{\infty} \lambda_k[\phi_k^{(\nu)}(t)]^2.$
One can then estimate derivatives by approximating $X_i^{(\nu)}$ with the first $K$ components 
\begin{equation}
\displaystyle X_{i,K}^{(\nu)}(t) = \mu^{(\nu)}(t) +\sum_{k=1}^{K}\xi_{ik}\phi_{k}^{(\nu)}(t).\label{eqn: X_1 KL approx}
\end{equation}
%and
%\begin{equation}
%\displaystyle \var(X_i^{(\nu)}(t)) \approx \sum_{k=1}^{K} \var(\xi_{ik})\{ \phi_k^{(\nu)}(t) \}^2,
%\end{equation}

In \eqref{eqn: X_1 KL} 
and  \eqref{eqn: X_1 KL approx}, $\mu^{(\nu)} $ is the $\nu$-th derivative of the mean function $\mu$ and can be estimated by local polynomial fitting applied to  a pooled scatterplot where one aggregates all the observed measurements from all sample trajectories. 
The FPCs $\xi_{ik}$ of the sample trajectories   can be estimated with the principal analysis by conditional expectation (PACE) approach described in \ci{mull:05:4}.  Starting from the eigenequations
% the $k$-th derivative of eigenfunctions $\phi_{k}^{(\nu)}(t)$, as eigenfunctions $\phi_{k}$ are the solutions of the eigen-equations
$\int G(s,t) \phi_k(s) ds = \lambda_k \phi_k (t)$ with orthonormality constraints, under suitable regularity conditions, by taking derivatives on both sides, one obtains targets and respective estimates,
\begin{equation*}%\label{eqn: phi1}
\phi_k^{(\nu)}(t)= \frac{1}{\lambda_k} \int \! G^{(0,\nu)}(s,t)\phi_k(s)ds, \quad 
\hat{\phi}_k^{(\nu)}(t)= \frac{1}{\hat{\lambda}_k} \int \! \hat{G}^{(0,\nu)}(s,t)\hat{\phi}_k(s)ds,
\end{equation*}
where $G^{(0,\nu)}(s,t) = {\partial^{\nu}}G(s,t)/{\partial t^{\nu}}$ is the $(0, \nu)$th partial derivative, $\hat{G}^{(0,\nu)}(s,t)$ is a smooth estimate of $G^{(0,\nu)}(s,t)$ obtained by for example local polynomial smoothing, and $\hat{\phi}_k$ an estimate of the $k$-th eigenfunction. The derivative $X_i^{(\nu)}$ is thus represented  by
\begin{equation*}
\hat{X}_{i,K}^{(\nu)}(t) = \hat{\mu}^{(\nu)}(t) +\sum_{k=1}^{K}\hat{\xi}_{ik}\hat{\phi}_{k}^{(\nu)}(t), % \label{eqn: X_1 KL est}
\end{equation*}
where the integer  $K$ is chosen in a data-adaptive fashion, for example by cross-validation \cp{rice:91},  AIC \cp{shib:81}, BIC \cp{schw:78}, and fraction of variance explained \cp{mull:09:1}. % Established methods for choosing $K$ include leave-one-curve-out \cp{rice:91},  AIC \cp{shib:81}, BIC \cp{schw:78}, and fraction of variance explained \cp{mull:09:1}.

A conceptual problem with this approach is that the eigenfunction derivatives $\phi_k^{(\nu)}$, $k=1,2,\ldots$ are not the orthogonal eigenfunctions of the derivatives $X_i^{(\nu)}$. Consequently this approach does not lead to the Karhunen--Lo\`eve expansion of derivatives, and therefore is suboptimal in terms of parsimoniousness. 
This motivates our next goal, to obtain the Karhunen--Lo\`eve representation for derivatives.

\subsection{Karhunen--Lo\`eve Representation for Derivatives} \label{ss:KL}

To obtain  the Karhunen--Lo\`eve representation  for derivatives,  consider the covariance function $G_{\nu}(s,t)=\cov(X^{(\nu)}(s), X^{(\nu)}(t))$ of $X^{(\nu)}$, $s,t \in \mathcal{T}$, a symmetric, positive definite and continuous function on $\mathcal{T} \times \mathcal{T}$. The associated autocovariance operator  $(A_{G_{\nu}}f)(t)=\int_{\mathcal{T}}G_{\nu}(s,t)f(s)ds$ for $f \in \mathcal{L}^2(\mathcal{T})$ is a linear Hilbert-Schmidt operator with eigenvalues denoted by $\lambda_{k,\nu}$ and orthogonal eigenfunctions  $\phi_{k,\nu}$, $k=1,2,\ldots$.  This leads to the representation
\begin{equation} \label{eqn:sdG11}
G_{\nu}(s,t)=  \sum_{k=1}^{\infty}\lambda_{k,\nu}\phi_{k,\nu}(s)\phi_{k,\nu}(t),
\end{equation}
with $\lambda_{1,\nu} > \lambda_{2,\nu} > \dots \ge 0$, and the Karhunen--Lo\`eve representation for the derivatives $X_i^{(\nu)}$,
\begin{equation}
X_i^{(\nu)}(t) = \mu^{(\nu)}(t) +\sum_{k=1}^{\infty}\xi_{ik,\nu}\phi_{k,\nu}(t), \hspace{0.15in} t \in \mathcal{T},\label{eqn: X_1 KLnew}
\end{equation}
with DPCs $\xi_{ik,\nu}=\int_{\mathcal{T}} \! (X^{(\nu)}_i(t)-\mu^{(\nu)}(t))\phi_{k,\nu}(t)dt$, for $i=1,\ldots, n$, $k\geq1$. For practical applications, one employs a truncated Karhunen--Lo\`eve representation
\begin{equation}\label{eq:XiK}
X_{i,K}^{(\nu)}(t) = \mu^{(\nu)}(t) +\sum_{k=1}^{K}\xi_{ik,\nu}\phi_{k,\nu}(t), 
\end{equation}
with a finite $K \ge 1$.

In contrast to \eqref{eqn: X_1 KL}, where derivatives of eigenfunctions are used in conjunction with the FPCs of processes $X$ to represent $X_i^{(\nu)}$, the proposed approach is based on the derivative principal component scores $\xi_{ik, \nu}$ (DPCs) and the derivative eigenfunctions $\phi_{k,\nu}$. The proposed representation \eqref{eq:XiK} is more efficient in representing $X_i^{(\nu)}$ than using \eqref{eqn: X_1 KL approx}, as 
  \begin{equation}\label{PCA-ineqn}
\sum_{k=1}^{K} \lambda_{k,\nu} \geqslant \sum_{k=1}^{K}\lambda_{k} \int \{\phi_k^{(\nu)}(t)\}^2dt, \hspace{0.15in} \text{for all } K\geq1.
  \end{equation}
Thus, for any finite integer $K\geq1$, the representation \eqref{eq:XiK} captures at least as much or more variation than  the representation \eqref{eqn: X_1 KL approx}. %This inequality is naturally rooted in the property of eigen-decompoistion.  The detailed discussion of it is  deferred to Appendix A.

The eigenfunctions of the derivatives  can be obtained by the spectral decomposition of $G_{\nu}$. Let $G^{(\nu,\nu)}={\partial^{2\nu}G}/{(\partial s^{\nu} \partial t^{\nu})}$. Under  regularity conditions,
\begin{align}
G_{\nu}(s,t) % &=E\{[X_i^{(\nu)}(s)-\mu^{(\nu)}(s)][X_i^{(\nu)}(t)-\mu^{(\nu)}(t)]\} \nonumber \\
&= E\left\{\frac{\partial^{\nu}}{\partial s^{\nu}}\frac{\partial^{\nu}}{\partial t^{\nu}}[X_i(s)-\mu(s)][X_i(t)-\mu(t)]\right\}  \nonumber \\
&= \frac{\partial^{\nu}}{\partial s^{\nu}}\frac{\partial^{\nu}}{\partial t^{\nu}} E\left\{[X_i(s)-\mu(s)][X_i(t)-\mu(t)]\right\}  \nonumber \\
&=G^{(\nu,\nu)}(s,t).  \label{eqn: G11}
\end{align}
%i.e. the covariance of derivatives is equal to a mixed partial derivative  of the covariance function $G(s,t)$.  
To fully implement our  approach, we need to identify the components of the representation \eqref{eqn: X_1 KLnew}, as  described in the next subsection.

\subsection{Sampling Model and BLUP} \label{ss:samp}

\bco
We assume that the longitudinal data are generated by a square integrable stochastic processes $X$ on a compact domain $\mathcal{T} \in \mathbb{R}$, with mean function denoted as $E\{X(t)\}=\mu(t)$ and auto-covariance function as $\text{cov}\{X(s),X(t)\}=G(s,t)$ for $s,t \in \mathcal{T}$. Our data model 
\fi

The sampling model
needs to reflect that longitudinal data are typically sparsely sampled with random locations of the design points, while functional data such as the spectral  data discussed in \autoref{ss:wheat} are sampled at a dense grid of design points. 
Assuming  that for the $i$-th trajectory $X_i, \,i=1, \ldots, n$, one obtains measurements  
$Y_{ij}$ made at random times $T_{ij} \in \cT$,  for $j=1,\ldots,N_i$, where for sparse longitudinal designs the number of observations per subject $N_i$ is bounded, while for dense functional designs $N_i = m \toinf$. For both scenarios the observed data are assumed to be generated as \begin{align} Y_{ij}= X_i(T_{ij})+ \epsilon_{ij}=\mu(T_{ij}) +\sum_{k=1}^{\infty}\xi_{ik}\phi_{k}(T_{ij})+\epsilon_{ij},\label{data_model} \end{align}
where $\epsilon_{ij}$ are i.i.d. measurement errors with $E (\epsilon_{ij})=0$ and $\var(\epsilon_{ij})= \sigma^2$, independent of $X_i$, and the $T_{ij}$ are generated according to some fixed density $f$ that has certain properties.  All expected values in the following are interpreted to be conditional on the random locations  $T_{ij}$, which is not explicitly indicated in the following.

Let $\SYi$ be an $N_i \times N_i$ matrix representing the covariance of ${\mathbf{Y}_i}$ with $(j,l)$-th element $(\Sigma_{\mathbf{Y}_i})_{j,l}=\text{cov}(Y_{ij},Y_{il})=G(T_{ij},T_{il})+\sigma^2 \delta_{jl}$, where $\delta_{jl} = 1$ if $j = l$ and 0 otherwise. In addition, $\bmui$ is a vector obtained by evaluating the mean function at the vector $(T_{i1},\ldots,T_{iN_i})$ of measurement times, and $\bzetaiknu$ is a column vector of length $N_i$ with  $j$-th element  cov$(\xi_{ik,\nu}, Y_{ij})$, $j=1,2,\ldots,N_i$, where 
\begin{align}\label{eqn: conditioning}
\cov (\xi_{ik,\nu},Y_{ij})
&=E\left[ \int \! (X_i^{(\nu)} (s)-\mu^{(\nu)}(s))\phi_{k,\nu}(s)ds \left\{X_i(T_{ij})-\mu(T_{ij})\right\}\right] \nonumber \\
&=  \int \!E \left[ (X_i^{(\nu)} (s)-\mu^{(\nu)}(s)) (X_i(T_{ij})-\mu(T_{ij})) \right] \phi_{k,\nu}(s)ds  \nonumber \\
&= \int \! G^{(\nu,0)}(s,T_{ij}) \phi_{k,\nu}(s)ds.
\end{align}
For the prediction of the DPCs  $\xiiknu$, we use the best linear unbiased predictors (BLUP, \cite{rice:01})
\begin{equation}
\txiiknu =  \bzetaiknu^T\SYiinv(\mathbf{Y}_i-\bmui) \label{eqn:FPC_1CE}
\end{equation}
that are always defined without distributional assumptions. In the special case that errors $\epsilon$ and processes $X$  are jointly Gaussian, $\txiiknu$ is the conditional expectation of $\xiiknu$ given $\bYi$, which is the optimal prediction of $\xiiknu$ under squared error loss.

\section{\sf Estimation of Derivative Principal Components } \label{s:est}
For estimation, we provide details for the most important case of the first derivative, $\nu=1$. Higher order derivatives are handled similarly.  By \eqref{eqn: X_1 KLnew}, approximate derivative representations are given by
  \begin{equation}
 X_{i,K}^{(1)}(t) = \mu^{(1)}(t) +\sum_{k=1}^{K}\xi_{ik,1}\phi_{k,1}(t), \label{eq:XiK1}
 \end{equation}
with approximation errors $\int (X_{i,K}^{(1)}(t)-X_i^{(1)}(t))^2 dt=\sum_{k=K+1}^{\infty}\lambda_{k,1}$, the convergence rate of which is determined by the decay rate of the $\lambda_{k,1}$.  We then obtain plug-in estimates for $ X_{i,K}^{(1)}$, $i=1,2,\ldots,n$, by substituting $ \mu^{(1)},  \phi_{k,1}$, and $
  \xi_{ik,1}$ in \eqref{eq:XiK1}  with corresponding estimates, leading to $ \hat{X}^{(1)}_{i,K}(t) = \hat{\mu}^{(1)}(t) +
\sum_{k=1}^{K}\hat{\xi}_{ik,1}\hat{\phi}_{k,1}(t)$.  Here we obtain $ \hat{\mu}^{(1)}(t)$
by applying local polynomial smoothing to a pooled scatterplot that aggregates the observed measurements from all sample trajectories, and  $\hlambda_{k, 1}$ and $\hat{\phi}_{k,1}(t)$ by spectral decomposition of $\hat{G}^{(1,1)}$, where $\hat{G}^{(1,1)}$ is the estimate for the mixed first-order partial derivative  of $G(s,t)$, obtained by two-dimensional local polynomial smoothing. For more details and related discussion about these estimates of $\mu^{(1)}(t)$ and $G^{(1,1)}(s,t)$ we refer to \autoref{app:meanEF}.

Estimating the DPCs $\xi_{ik,1} $  is an essential step for representing  derivatives as in  \eqref{eq:XiK1}.  From the definition $\xi_{ik,\nu}=\int (X^{(\nu)}_i(t)-\mu^{(\nu)}(t))\phi_{k,\nu}(t)dt$, it seems plausible to obtain $\hat{\xi}_{ik,1}$ using  plug-in estimates and numerical integration, $ \hat{\xi}_{ik,1} = \int \! \{ \hat{X}_{i,K}^{(1)}(t)- \hat{\mu}^{(1)}(t)\} \hat{\phi}_{k,1}(t) dt$.  However, this approach requires that one already has derivative estimates $\hat{X}_{i,K}^{(1)}(t)$, which is not viable, especially for sparse/longitudinal designs. 

An alternative approach is  to construct BLUPs $\txiiknu$ from the observed measurements $\mathbf{Y}_i$ that were made at  time points $\mathbf{T_i}=(T_{i1}, T_{i2},\ldots,T_{iN_i})^T$ as in \eqref{eqn:FPC_1CE}, where $\txiiknu$ can be consistently estimated.
Applying  \eqref{eqn:FPC_1CE} for $\nu=1$, the BLUP for $\xi_{ik,1}$ given  observations $\mathbf{Y}_i$  is
\begin{equation}
  \tilde{\xi}_{ik,1}= \bzetaik^T \SYiinv(\bYi - \bmui),  \label{eqn:xi_ik1}
\end{equation}
where $\bzetaik$ is the covariance vector of $\xi_{ik,1}$ and $\mathbf{Y}_i$, with length $N_i$ and $j$-th element $\zeta_{ikj}=\int \! G^{(1,0)}(s,T_{ij}) \phi_{k,1}(s)ds$, as per \eqref{eqn: conditioning}.
Estimates $\hxiikone$ for the $\txiikone$ are then obtained by substituting estimates for $\bzetaik$, $\SYi$, and $\bmui$ in \eqref{eqn:xi_ik1},
\begin{equation} \label{eq:hxiikone}
\hxiikone = \hbzetaik^T \hSYiinv (\bYi - \hbmui),
\end{equation}
where $\hat{\zeta}_{ikj}=\int \! \hat{G}^{(1,0)}(s,T_{ij}) \hat{\phi}_{k,1}(s)ds$ and $(\hat{\Sigma}_{\mathbf{Y}_i})_{j,l}=\hat{G}(T_{ij},T_{il})+\hat{\sigma}^2 \delta_{jl}$.

When the joint Gaussianity of $\epsilon$ and $X$ holds, $\tilde{\xi}_{ik,1}$ is the conditional expectation of $\xi_{ik,1}$ given $\bYi$, the best prediction.  The required estimates  $\hat{G}^{(1,0)}(s,T_{ij})$ of the partial derivative of the covariance function and estimates $\hat{G}(T_{ij},T_{il})$ can be obtained  by local polynomial smoothing (\cite{mull:09:1} equation (7)). Estimate $\hat{\sigma}^2$ of the error variance $\sigma^2$  can be obtained using the method  described in equation (2) of \ci{mull:05:4}. 

In practice, the number of included components $K$ can be chosen by a variety of methods, including leave-one-curve-out cross-validation \cp{rice:91},  pseudo-AIC \cp{shib:81}, or pseudo-BIC \cp{schw:78, mull:05:4}. Another fast and stable option that works quite well in practice is to choose the smallest $K$ so that the inclusion of the first $K$  components explains  a preset level of variation, which can be set at 90\%. % For regression models, where functional derivatives serve as predictors of responses, another option is to view the derivative principal components as nested sets of predictors and to use regression-AIC or BIC to select $K$,. In a prediction setting, direct minimization of a cross-validation prediction error $CV(L)=\sum[y_i-f(\xi_{i11}^{(-i)},\ldots,\xi_{iK1}^{(-i)})]^2$, often implemented as $L-$fold cross-validation, often is a good option.  Here $f(\cdot)$ represents the selected regression model and $\xi_{ik1}^{(-i)}$ are the estimates constructed while omitting the data for the $i$-th subject.

\section{\sf Asymptotic Results} \label{s:asymp}
We take a unified approach in our estimation procedure for the DPCs and other model components that encompasses both the dense and the sparse case. Estimation of the derivatives of the mean and the covariance function, and the derivative eigenfunctions are based on smoothing the pooled scatter plots \citep{zhan:16}; the estimation for the DPCs is based on best linear unbiased predictors as in \eqref{eq:hxiikone}. 
%Best linear predictors are equivalent to conditional expectations if Gaussian assumptions are satisfied.
We derive  convergence rate results that make use of a novel argument for the dense case. Consistency of the estimator $\hat{X}_{i,K}^{(1)}$ for $X_{i,K}^{(1)}$ can be obtained by utilizing the convergence of  estimators $\hmuone(t)$, $\hat{G}^{(1,1)}(s,t)$, and $\hxiikone$ to their respective targets $\muone(t)$, $G^{(1,1)}(s,t)$, and $\xiikone$ as in Theorems~\ref{thm:muG} and \ref{thm:xi} below. %\autoref{thm:muG} and \autoref{thm:xi} below. 
Regularity conditions include assumptions on the number and distribution of the design points, smoothness of the mean and the covariance functions, bandwidth choices, and moments for $X(t)$, as detailed in \autoref{app:proofs}. 

We present results on asymptotic convergence rates in the supremum norm for the estimates of the mean and the covariance functions for derivatives and the corresponding estimates of eigenfunctions. Our first theorem covers the case of sparse/longitudinal designs, where the number of design points $N_i$ is bounded, and the case of dense/functional designs, where $N_i = m \toinf$. For convenience of notation, let
\begin{equation}
a_{n1}=h_\mu^2 + \sqrt{\frac{\log(n)}{nh_\mu}}, \quad b_{n1} = h_G^2 + \sqrt{\frac{\log(n)}{nh_G^2}},
\end{equation}
%and for the dense case 
\begin{equation}
a_{n2}=h_\mu^2 + \sqrt{\left(1 + \frac{1}{mh_\mu}\right)\frac{\log(n)}{n}} , \quad b_{n2} = h_G^2 + \left(1 + \frac{1}{mh_G}\right)\sqrt{\frac{\log(n)}{n}}.
\end{equation}

\begin{thm} \label{thm:muG} Suppose \autoref{a:K}--\autoref{a:Ubeta} in \autoref{app:proofs} hold. Setting $a_n = a_{n1}$ and $b_n = b_{n1}$ for the sparse case when $N_i \le N_0 < \infty$, and $a_n = a_{n2}$ and $b_n = b_{n2}$ for the dense case when $N_i = m \toinf$, for $i = 1, \dots, n$, 
	\begin{align}
	\supt |\hmuone(t) - \muone(t)| & = O(a_{n}) \quad \almostsure, \label{eq:mu1Rt}\\
	\supst |\hGone(s,t) - \Gone(s,t)| & = O(a_{n} + b_{n}) \quad \almostsure, \label{eq:G11Rt}\\
	\supt |\hphikone(t) - \phikone(t)| & = O(a_{n} + b_{n}) \quad \almostsure \label{eq:phi1Rt}
	\end{align}
	for any $k \ge 1$.
%\item (Dense) When $N_i = m \toinf$ for all $i = 1, \dots, n$, 
%	\begin{align} 
%	\supt |\hmuone(t) - \muone(t)| & = O(b_{n1}) \quad \almostsure, \label{eq:muDeRt}\\
%	\supst |\hGone(s,t) - \Gone(s,t)| & = O(b_{n1} + b_{n2}) \quad \almostsure, \label{eq:GDeRt}\\
%	\supt |\hphikone(t) - \phikone(t)| & = O(b_{n1} + b_{n2}) \quad \almostsure, \label{eq:phiDeRt}
%	\end{align}
%	for any $k \ge 1$.
\end{thm}
This result provides the basis for the convergence of the DPCs. We write $\alpha_n \asymp \beta_n$ if $K_1 \alpha_n \le \beta_n \le K_2 \alpha_n$ for some constants $0 < K_1 < K_2 < \infty$. In the sparse case, the optimal supremum convergence rates for $\hGone$, and $\hphikone$ are of order $O((n / \log(n))^{-1/3})$ almost surely, achieved for example if $h_\mu \asymp (n  / \log(n))^{-1/5}$, $h_{G} \asymp (n / \log(n))^{-1/6}$, $\alpha > 5/2$, and $\beta > 3$ as in \autoref{a:Ualpha} and \autoref{a:Ubeta}. In the dense case, if the number of observations per curve $m$ is at least of order $(n / \log(n))^{1/4}$, then a root-$n$ rate is achieved for our estimates if $h_\mu,h_G \asymp (n/\log(n))^{-1/4}$, $\alpha > 4$, and $\beta > 4$. 

Using asymptotic results in \ci{mull:09:1} for auxiliary estimates of the mean and the covariance functions and their derivatives or partial derivatives, we obtain asymptotic convergence rates of $\hat{\xi}_{ik,1}$ toward the appropriate target, $\tilde{\xi}_{ik,1}$, as in \eqref{eqn:xi_ik1} for the sparse case and $\xi_{ik,1}$ in the dense case.
%proof: Following the uniform convergence of $\hat{G}_{1,0}(t)$ for $G_{1,0}(t)$ proved in \autoref{thm:muG} of Bitao (08),

\begin{thm} \label{thm:xi}
Under the conditions of \autoref{thm:muG}, 
\begin{equation}
|\hxiikone - \txiikone| = O_p(N_i^2 (a_n + b_n)).
\end{equation}
If furthermore \autoref{a:phi1Finite} holds, $(X, \epsilon)$ are jointly Gaussian, and $N_i = m \toinf$, then
\begin{equation}
|\txiikone - \xiikone| = O_p(m^{-1/2}).
\end{equation}
\end{thm}

%The second rate for the BLUP $\txiikone$ is new, argued based on quadrature approximation to integral operators. 
For example, in the sparse case if we choose $h_\mu \asymp (n/\log(n))^{-1/5}$ and $h_G\asymp (n / \log(n))^{-1/6}$, then $|\hat{\xi}_{ik,1}-\tilde{\xi}_{ik,1}|=O_p((n/\log(n))^{-2/5})$. In the dense case, the $\xiikone$ can be consistently estimated if $m h_\mu \tozero$, $m h_G \tozero$, and $m = o((n/\log(n))^{1/4})$, with the optimal rate for $|\hxiikone - \xiikone| = O_p((n/\log(n))^{-1/3} m^{4/3} + m^{-1/2})$, achieved when $h_\mu, h_G = (n/\log(n))^{-1/6} m^{-1/3}$. 

Here, we define $\hXone_{i,K}$ similarly to $\Xone_{i,K}$ in \eqref{eq:XiK1}, except that we replace the population quantities by their corresponding estimates, and $\tX_{i,K}^{(1)}$ by replacing $\xi_{ik,1}$ with $\txi_{ik,1}$ in \eqref{eq:XiK1}. 

\begin{thm} \label{thm:X} Assume the conditions of \autoref{thm:muG} hold. For all $i=1,2,\dots,n$, and any fixed  integer $K$,
\begin{equation}
\normSup{\hat{X}^{(1)}_{i,K}(t)-\tilde{X}^{(1)}_{i,K}(t)} = O_p(N_i^2 (a_n + b_n)).
\end{equation}
If furthermore \autoref{a:phi1Finite} holds, $(X, \epsilon)$ are jointly Gaussian, and $N_i = m \toinf$, then
\begin{equation}
\normSup{\hat{X}^{(1)}_{i,K}(t)-\Xone_{i,K}(t)} = O_p(m^2 (a_n + b_n) + m^{-1/2}).
\end{equation}
\end{thm}

If we choose the bandwidths as described after \autoref{thm:xi}, then in the sparse case  $\normSup{\hat{X}^{(1)}_{i,K}(t)-\tilde{X}^{(1)}_{i,K}(t)} = O_p((n/\log(n))^{-2/5})$, and in the dense case $\normSup{\hat{X}^{(1)}_{i,K}(t)-\Xone_{i,K}(t)} = O_p((n/\log(n))^{-1/3} m^{4/3} + m^{-1/2})$.

%Our next result pertains to the sparse case only. Let $\bxi_{iK,1} = (\xi_{i1,1}, \dots, \xi_{iK,1})^T$, $\bH = (\bzeta_{i1}, \dots, \bzeta_{iK})$ be an $N_i \times K$ matrix, $\bLambda = \diag(\lambda_{1,1}, \dots, \lambda_{K,1})$, $\tbxi_{iK,1} = (\txi_{i1,1}, \dots, \txi_{iK,1})^T$, $\hbH = (\hbzeta_{i1}, \dots, \hbzeta_{iK})$ and $\hbLambda = \diag(\hlambda_{1,1}, \dots, \hlambda_{K,1})$. 
%\begin{thm}\label{thm:norm} 
%Assume \autoref{a:K}--\autoref{a:Ubeta}, \autoref{a:omega}, $\supi N_i \le N_0 < \infty$, and that $\xi_{ik}$ and $\epsilon_{ij}$ are jointly Gaussian. Then for all $t \in \T$ and  $i=1,2,\ldots,n$, 
%\begin{align} \label{thmXiKnorm-eq}
%\lim_{K \rightarrow \infty} \lim_{n \rightarrow \infty}P|\frac{\hat{X}^{(1)}_{i,K}(t)-X_i^{(1)}(t)}{\sqrt{\hat{\omega}_K(t,t)}}<x |=\Phi(x),
%	\end{align}
%	where $\hat{\omega}_K(s,t)=\hbphi_{K,1,s}^T\hbOmega_K \hbphi_{K,1,t}$, $\hbphi_{K,1,t}=(\hat{\phi}_{1,1}(t),\hat{\phi}_{2,1}(t),\ldots, \hat{\phi}_{K,1}(t))^T$, $\hbOmega_K=\widehat{\textup{var}}(\tbxi_{iK,1}-\bxi_{iK,1}) = \hbLambda-\hbH^T\SYiinv\hbH$, and $\Phi$ is the standard Gaussian c.d.f.
%\end{thm}
%
%Using this result, we can construct approximate $(1-\alpha)100\%$ pointwise confidence bands for derivatives $\Xone_{i}(t)$ by $\hat{X}^{(1)}_{i,K}(t) \pm\sqrt{\hat{\omega}_K(t,t)}\Phi(1-\frac{\alpha}{2})$.

%\bc {\bf \sf 4. \quad SIMULATION STUDIES}\sm \ec \rs\no
\section{\sf Simulation Studies} \label{s:sim}
To examine the practical utility of the DPCs, we compared  them with various alternatives under different simulation settings, which included a dense and a sparse design.
To evaluate the performance of each method in terms of recovering the true derivative trajectories, we examined the mean and standard deviation of the relative mean integrated square errors (RMISE), defined as
\begin{align} \label{RMSE}
\displaystyle \text{RMISE}=\frac{1}{n}\sum_{i=1}^n\frac{ \int_{0}^1 \{\hat{X}_i^{(1)}(t)- X_i^{(1)}(t)\}^2dt}{\int_0^1\{X_i^{(1)}(t)\}^2dt }.
\end{align}
We compared the proposed approach based on model \eqref{eqn: X_1 KLnew}, referred to as DPCA, and a PACE method \citep{mull:05:4} followed by differentiating  the eigenfunctions of observed processes as in \ci{mull:09:1}, corresponding to  \eqref{eqn: X_1 KL}, referred to as FPCA.
Each simulation consisted of 400 Monte Carlo samples with the number of random trajectories chosen as $n=200$ per simulation sample.

While our methodology is intended to address the difficult problem of derivative estimation for the case of sparse designs, the Karhunen--Lo\`eve expansion for derivatives is of interest in itself and is also applicable to densely sampled functional data. The proposed method also has advantages for densely sampled data with large measurement errors. For the case of dense designs, another straightforward approach to obtain derivatives is  LOCAL, a method that corresponds to local quadratic smoothing of each trajectory separately, then taking the coefficient at the linear term as estimate of the derivative; and SMOOTH-DQ, where difference quotients are smoothed with local linear smoothers.  These methods are obvious tools to obtain derivatives, but their application is  only reasonable for densely sampled trajectories.

All simulated longitudinal data  were generated according to the data sampling  model described in Section 2, with mean function $\displaystyle \mu(t) =4t+(0.02\pi)^{-1/2}\exp \{-(t-0.5)^2/[2(0.1)^2]\}$;
five eigenfunctions $\phi_k$, where $\phi_k$ is the $k$th orthonormal Legendre polynomial on $[0, 1]$; eigenvalues
$\lambda_k =3,\, 2,\, 1,\, 0.1,\, 0.1$ for $k=1, \ldots,5$; and FPC scores $\xi_{ik}$ distributed as $\mathcal{N}(0,\lambda_k)$, $k=1,2,\ldots,5$. The additional measurement errors $\epsilon_{ij}$ were i.i.d $\mathcal{N}(0,\sigma^2)$, where the value of $\sigma$ varied for different simulation settings. % The following two simulation scenarios were considered: %with \textit{Simulation A} under small noise level settings closest to  the wheat spectra data and \textit{Simulation B}  resembling systolic blood pressure longitudinal measurements in the BLSA study:

\textit{Simulation A -- Sparsely Sampled Longitudinal Data.}
The number  of observations for each trajectory, denoted by $N_i$, was generated from a discrete uniform distribution from 2 to 9. The measurement times of the observations were randomly sampled in $[0,1]$ according to a Beta($2/3, 1$) distribution with mean 0.4 and standard deviation 0.3, so that the design is 
genuinely sparse and unbalanced.
Measurement errors were generated by a Gaussian distribution with standard deviation $\sigma = 0.5$ or $\sigma = 1$.

\textit{Simulation B -- Densely Sampled Functional Data.} 
Each random trajectory consists of 51 equidistant observations measured at the same dense time grid on the interval $ [0,1]$. In this setting, the proposed DPC method is compared with FPC, LOCAL, and SMOOTH-DQ.   In LOCAL, we estimate the derivatives  by applying local quadratic smoothing to individual subjects, with bandwidth selected by minimizing the average cross-validated integrated squared deviation between the resulting derivatives and the raw difference quotients formed from adjacent measurements. In SMOOTH-DQ, individual derivative trajectories were estimated by local linear smoothing of  the difference quotients, with smoothing bandwidth chosen by a similar strategy as for LOCAL.  Gaussian measurement errors were added with standard deviation $\sigma=1$  or $\sigma = 2$.

For the smoothing steps, Gaussian kernels were used and the bandwidths $h_{\mu}$, $h_{G}$  were selected by a generalized cross-validation method (GCV). For DPC, we took the partial derivative of $\hG^{(0,0)}$ to obtain $\hG^{(1,0)}$, which was superior in performance compared to  smoothing the raw data directly, and then we applied a one-dimensional smoother on $\hG^{(1, 0)}$ to obtain $\hG^{(1,1)}$, where the smoothing bandwidth was chosen to be the same as $h_{G}$. The smoothers for $\hG^{(1,0)}$ and $\hG^{(1,1)}$ enjoy better finite sample performance than two-dimensional smoothers due to more stable estimates and better boundary behavior. We let the number of components $K$ range from 1 to 5 for estimating the derivative curves, and we also included an automatic selection of $K$ based on FVE with threshold 90\%. The population fraction of variance explained for FPCA is $\sum_{k=1}^{K}\lambda_{k} \int \{\phi_k^{(\nu)}(t)\}^2dt / \sum_{k=1}^5 \lambda_{k,\nu}$, which were 0\%, 18\%, 61\%, 74\%, 100\%  for $K=1, \dots, 5$, respectively. In contrast, the FVEs for  DPCA are $\sum_{k=1}^{K} \lambda_{k,\nu} / \sum_{k=1}^5 \lambda_{k, \nu}$, which were 56\%, 77\%, 92\%, 100\%, 100\% in our simulation. It is evident that DPCA explains more variance than FPCA, given the same number of components, as expected in view of  \eqref{PCA-ineqn}.

The results for sparse and irregular designs (\textit{Simulation A}) are shown in \autoref{table:sparse}. 
For sparse and irregular designs, the sparsity of the observations for each subject precludes the applicability of LOCAL and SMOOTH-DQ, so we compared the proposed DPCA only with FPCA, given that the latter was shown  to have much better performance compared to mixed effect modeling with B-splines in \cite{mull:09:1}.  We also include the RMISE for the simple approach of estimating individual derivatives by the  estimated population mean derivative $\hat{\mu}^{(1)}$.  

\begin{table}
\centering
\single\caption{RMISE for \textit{Simulation A}, sparse designs, with error standard deviations $\sigma=0.5$ or $\sigma=1$. We report the mean of the RMISE based on 400 Monte Carlo repeats, where the standard deviations are all between 0.07 and 0.09 (not shown). The first 5 columns correspond to FPCA and DPCA using different fixed numbers of components $K$; the 6th column corresponds to selecting $K$ according to FVE, with the mean of the selected $K$ in brackets. } 
\label{table:sparse}
\begin{tabular}{|c|ccccc|c|c|}
	\hline
	$\sigma=0.5$ &     $K=1$     &     $K=2$     &      $K=3$     &      $K=4$      &      $K=5$      & FVE &       $\hat\mu^{(1)}$        \\
	\hline
	    FPCA      & 0.59 & 0.53  &  0.44  & {0.43 } &  0.44  & 0.44 (4.6) & \multirow{2}{*}{0.59} \\
	    DPCA      & 0.50  & 0.44 & {0.43} & {0.43} & {0.43} & 0.43 (2.2) &  \\
	\hline\hline
	 $\sigma=1$  &   $K=1$     &     $K=2$     &      $K=3$     &      $K=4$      &      $K=5$      & FVE &       $\hat\mu^{(1)}$     \\
	\hline
	    FPCA      & 0.60  & 0.54 & {0.46} & {0.46} & {0.46} & 0.46 (4.5) & \multirow{2}{*}{0.59} \\
	    DPCA      & 0.52 & 0.47 & {0.46} & {0.46} & {0.46} & 0.46 (2.2) &  \\
	\hline
\end{tabular}
\end{table}

\begin{figure}[H]
\begin{center}
\includegraphics[width=5in]{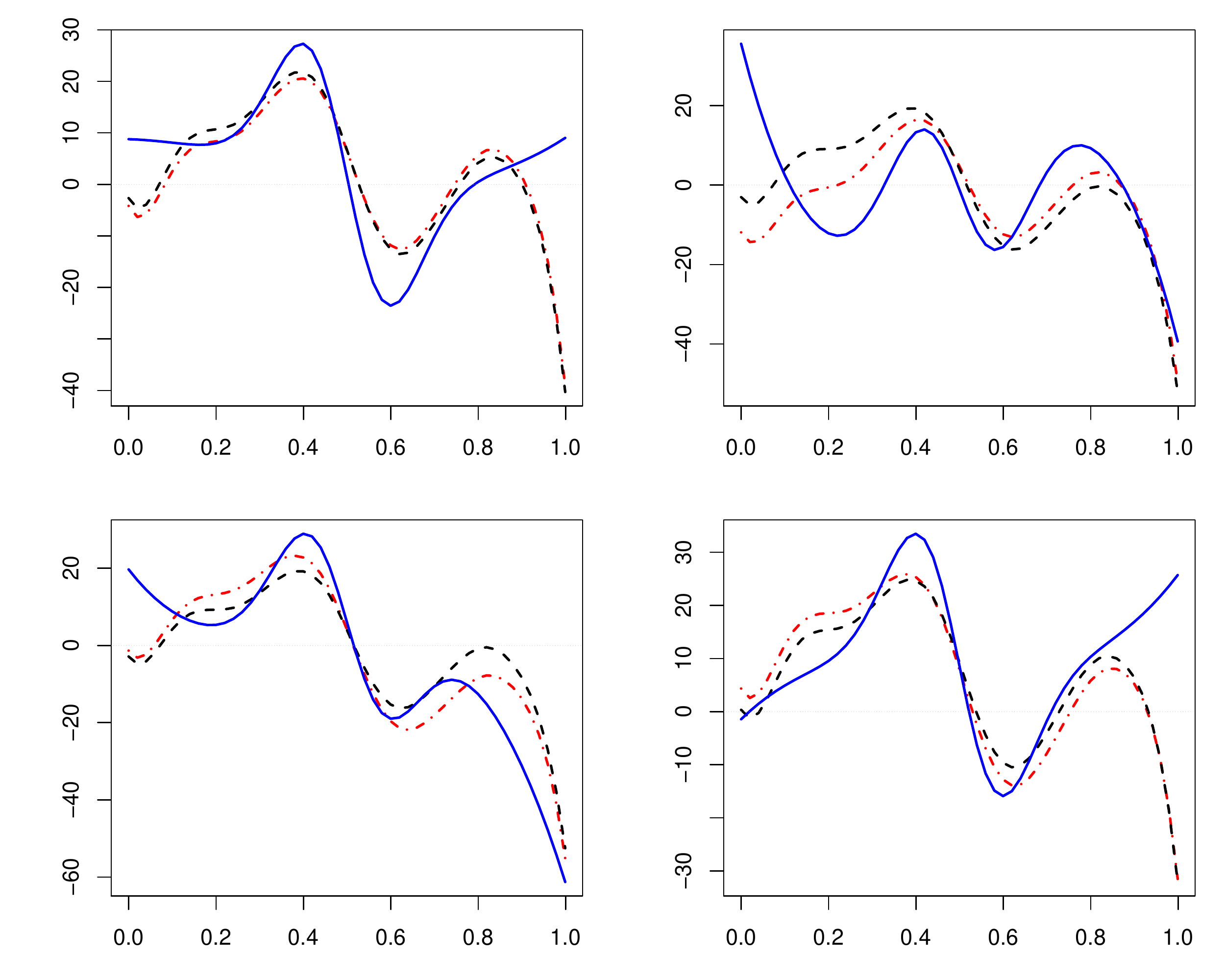}
\single \caption{\small{True derivative curves and the corresponding estimates obtained by FPCA and DPCA, for four randomly selected sparsely sampled trajectories generated in  \textit{Simulation A} (sparse designs) with $\sigma=2$.  Each of the four panels represents an individual derivative trajectory and consists of the true underlying derivative (solid), the derivative estimates by FPCA (dashed) and by DPCA (dash-dot). }}
\label{figure_sparse}
\end{center}
\end{figure}

As the results in \autoref{table:sparse} demonstrate, given the same number of components $K$, the representation of derivatives with DPCA works equally well or better than  FPCA in terms of RMISE where, in the latter, derivatives  are represented with the standard FPCs and the derivatives of the eigenfunctions. DPCA performs well with as few as $K=2$ components, while FPCA  performs well only when $K \ge 3$. The performance for individual trajectories when  $K=2$ is illustrated in \autoref{figure_sparse}, which shows the derivative curves and corresponding estimates obtained with FPCA and DPCA  for four randomly selected samples generated under measurement error $\sigma=1$. We find that in the sparse case the estimated derivatives using FPCA and DPCA are overall similar. % especially when a large $K$ is used.

%\vspace{-0.5in}

The results for \textit{Simulation B} for dense designs are shown in \autoref{table:dense}. We found that under both small ($\sigma=1$) and large ($\sigma=2$) measurement errors, the proposed DPCA clearly outperforms the other three methods in terms of RMISE.  The runner-up among the other methods is FPCA, but it was highly unstable for more than  five components.  Performance for all methods was better  with smaller measurement errors ($\sigma=1$), due to the fact that it is particularly difficult to infer derivatives in situations with large measurement errors. Also, unsurprisingly,  under the same level of measurement errors, all methods achieve smaller RMISE for dense designs, compared to their respective performance under sparse designs. This shows that DPCA has a significant advantage over FPCA in the dense setting. 

 \vspace{0.3in}
\begin{table}[H]
\centering
\single\caption{Relative mean integrated squared errors (RMISE) for \textit{Simulation B}, dense designs, with error standard deviation $\sigma=1$ or $\sigma=2$. We report the mean of the RMISE based on 400 Monte Carlo repeats, where the standard deviations are all between 0.01 and 0.02 for all except LOCAL. For LOCAL, the derivative of each curve is estimated individually using local quadratic kernel smoothing; for SMOOTH-DQ, the derivative of each curve is obtained via smoothing of the difference quotients of the observed measurements. The first 5 columns correspond to FPCA and DPCA using different numbers of fixed $K$; the 6th column corresponds to selecting $K$ according to FVE, with the mean  of the selected $K$ in brackets.} \vspace{.7cm}
\label{table:dense}
\begin{tabular}{|c|ccccc|c|cc|}
	\hline
	$\sigma=1$ &  $K=1$     &     $K=2$     &      $K=3$     &      $K=4$      &      $K=5$ & FVE &            LOCAL             &          SMOOTH-DQ           \\
	\hline
	   FPCA     & 0.51 & 0.42 & 0.27 & 0.2  &     0.16      & 0.16 (5.0) &    \multirow{2}{*}{0.23}     & \multirow{2}{*}{0.65} \\
	   DPCA     & 0.32 & 0.22 & 0.16 & 0.13 & \textbf{0.08} & 0.13 (3.9) &                              &  \\
	\hline\hline
	$\sigma=2$ & $K=1$     &     $K=2$     &      $K=3$     &      $K=4$      &      $K=5$   & FVE &            LOCAL             &          SMOOTH-DQ           \\
	\hline
	   FPCA     & 0.51 & 0.43 & 0.29 & 0.27 &     0.26      & 0.26 (5.0) & \multirow{2}{*}{0.51} & \multirow{2}{*}{0.76} \\
	   DPCA     & 0.34 & 0.26 & 0.22 & 0.19 & \textbf{0.18} & 0.20 (3.9) &                              &  \\
	\hline
\end{tabular}
\end{table}
 
 \vspace{.7cm}

\section{\sf Applications} \label{s:application}
\subsection{\sl Modeling Derivatives of Tammar Wallaby Body Length Data} \label{ss:wallaby}
We applied the proposed DPCA for derivative estimation to the Wallaby growth data, which can be found at  http://www.statsci.org/data/oz/wallaby.html, from the Australian Dataset and Story Library (OzDASL). This dataset includes body length measurements for 36 tammar wallabies (\textit{Macropus eugenii}),  longitudinally taken and collected from wallabies in their early age. A detailed introduction of the dataset is given by \ci{mall:94}. To gain a better understanding of the growth pattern of wallabies, we investigated the dynamics of their  body length growth  by estimating the derivatives of their growth trajectories. 

\begin{figure}[H]
\begin{center}
\vspace{0.1in}
\includegraphics[width=5in]{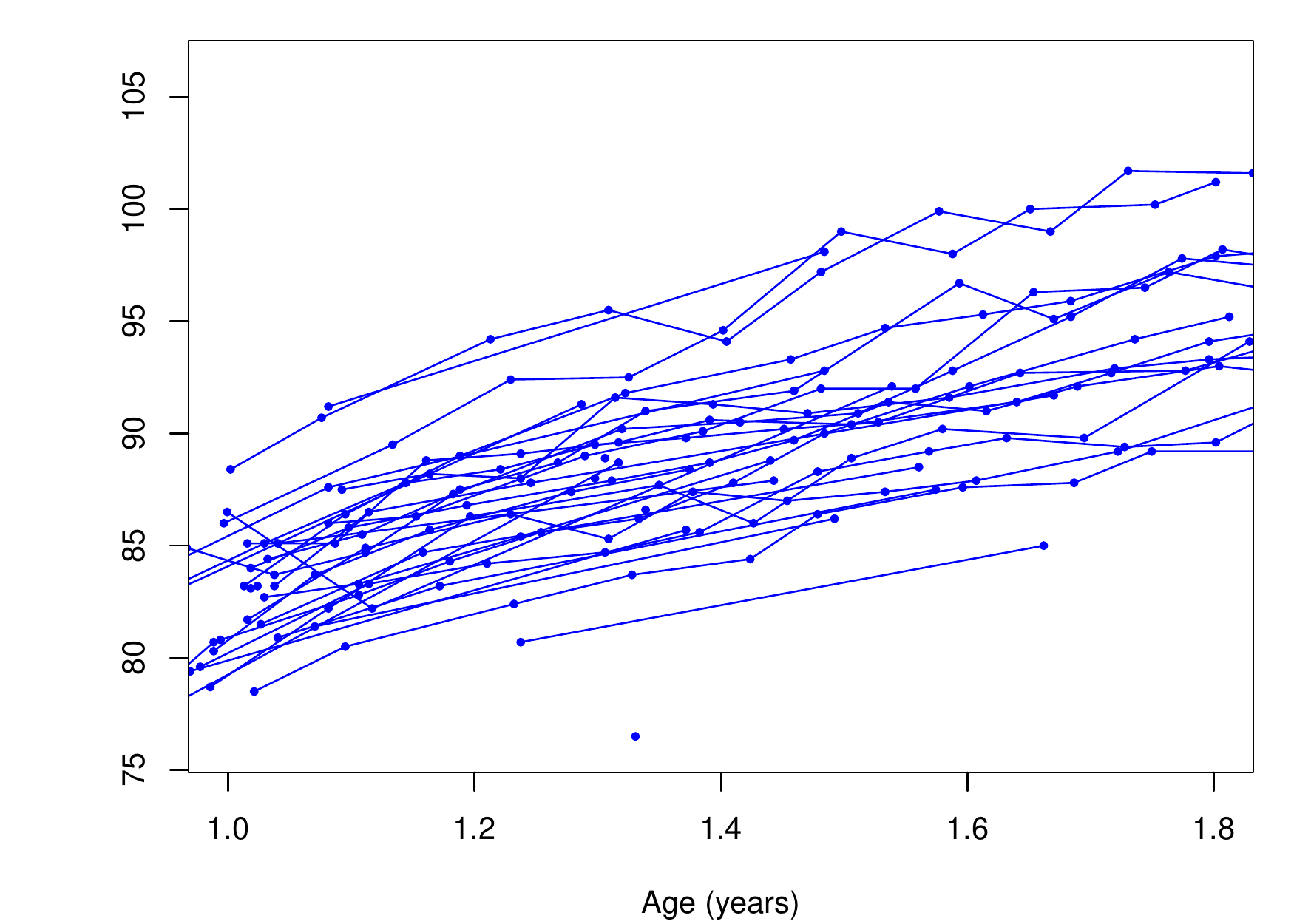}
\single \caption{\small{The trajectory spaghetti plot of body length growth for 36 wallabies. The recorded measurements over time for each wallaby are quite sparse, with measurement counts ranging from 1 to 12.}}
\label{Wallaby_trajectory}
\end{center}
\end{figure}

One main difficulty is that the body length measurements are very sparse, irregular, and fragmentary, as shown in \autoref{Wallaby_trajectory}, making these data  a good test case to reveal the difficulties in recovering  derivatives from sparse longitudinal designs. The 36 wallabies included in the dataset had their body length measured from 1 to 12 times per subject, with a median of 3.5 measurements per subject.

\begin{figure}[H]
\begin{center}
\includegraphics[width=6in]{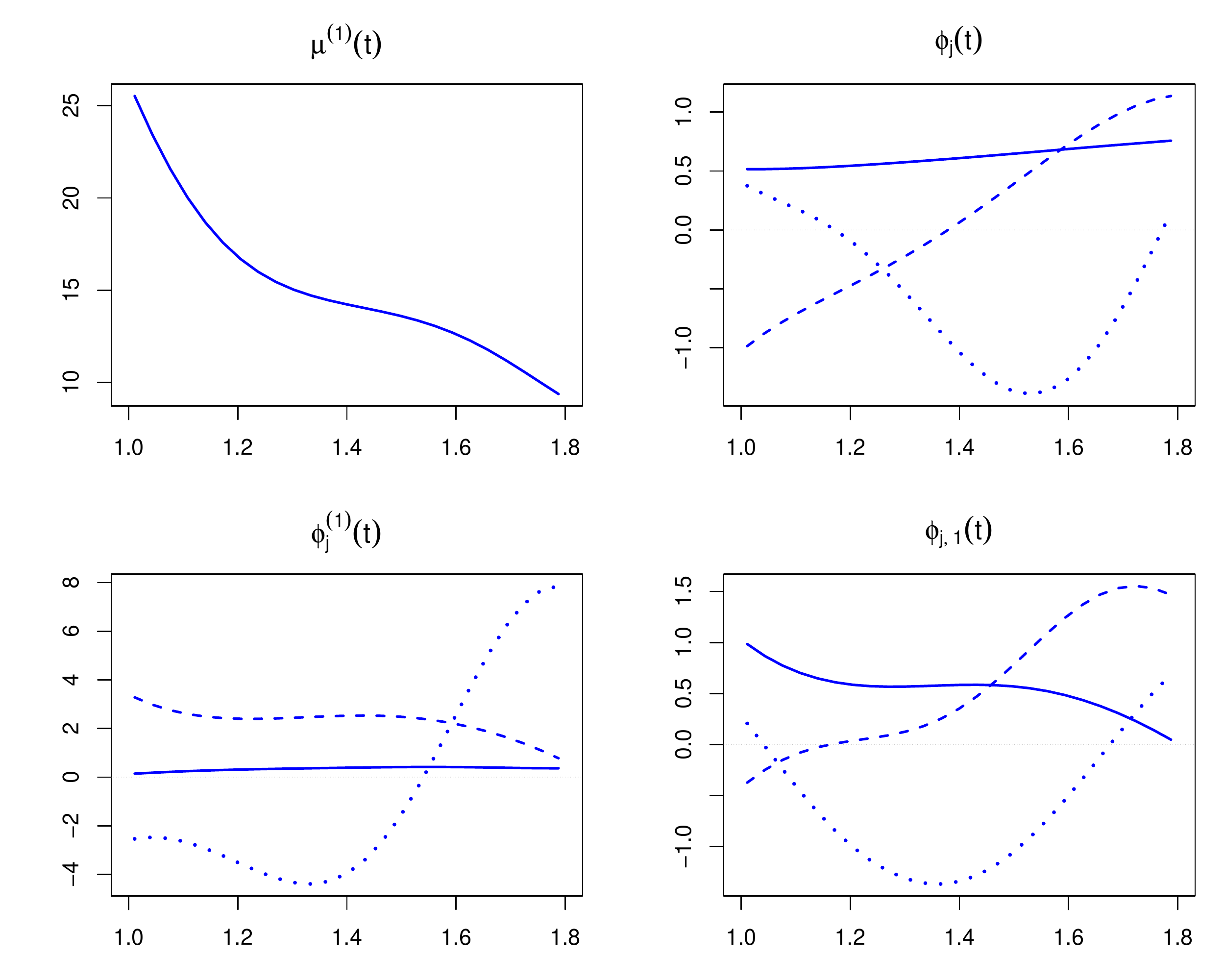}
\single\caption{\small{Upper left: estimated mean derivative function of the body length for wallabies; upper right: estimated first three eigenfunctions of body length trajectories from FPCA, explaining  $97\%$, $2.4\%$, and 0.59\% of overall variance in the trajectories, respectively; lower left: estimated derivatives of the eigenfunctions of body length trajectories in the upper right panel; lower right: estimated eigenfunctions for derivatives from DPCA, explaining 62.8\%, 26.1\%, and 10.9\% of overall variance in the derivatives. First, second and third  eigenfunctions are denoted by solid, dashed, and dotted lines, respectively. }}
\label{Wallaby_eigen}
\end{center}
\end{figure}

Aiming at a number of components with 90\% FVE leads to inclusion of  the first $K=3$ derivative eigenfunctions. The estimated first derivative of the mean function, eigenfunctions of the original growth trajectories, 
and those of the derivatives by the proposed approach are shown  in \autoref{Wallaby_eigen}. The average  dynamic changes in body length are represented by the mean derivative function (upper left panel), which exhibits a monotonically decreasing trend, from greater than 25 cm/yr at age 1 to less than 10 cm/yr at age 1.8, where the decline rate  of the mean derivative function becomes generally slower as age increases. The first eigenfunction (solid) of the trajectories reflects overall body length at different ages, the second eigenfunction (dashed) characterizes a contrast in length  between early and late stages, and the third eigenfunction (dotted) corresponds to a contrast between a period around 1.5 years and the other stages (upper right panel). 

The primary mode of dynamic changes in body length, as reflected by  the first eigenfunction of the derivatives in the lower right panel (solid) of \autoref{Wallaby_eigen}, represents the overall speed of growth which has a %heteroskedasticity component and is
 decreasing trend as wallabies get older. The second mode of dynamic variation is determined by the second eigenfunction of the derivatives (dashed) that mainly contrasts dynamic variation during young age with that of late ages. The third eigenfunction of the derivatives (dotted) emphasizes growth variation around age 1.38 and stands for a contrast of growth speed between middle age and early/late ages. The  eigenfunctions of derivatives are seen to clearly differ from the derivatives of the eigenfunctions (lower left panel) of the trajectories themselves, and are well interpretable.

\begin{figure}[H]
\begin{center}
\includegraphics[width=6in]{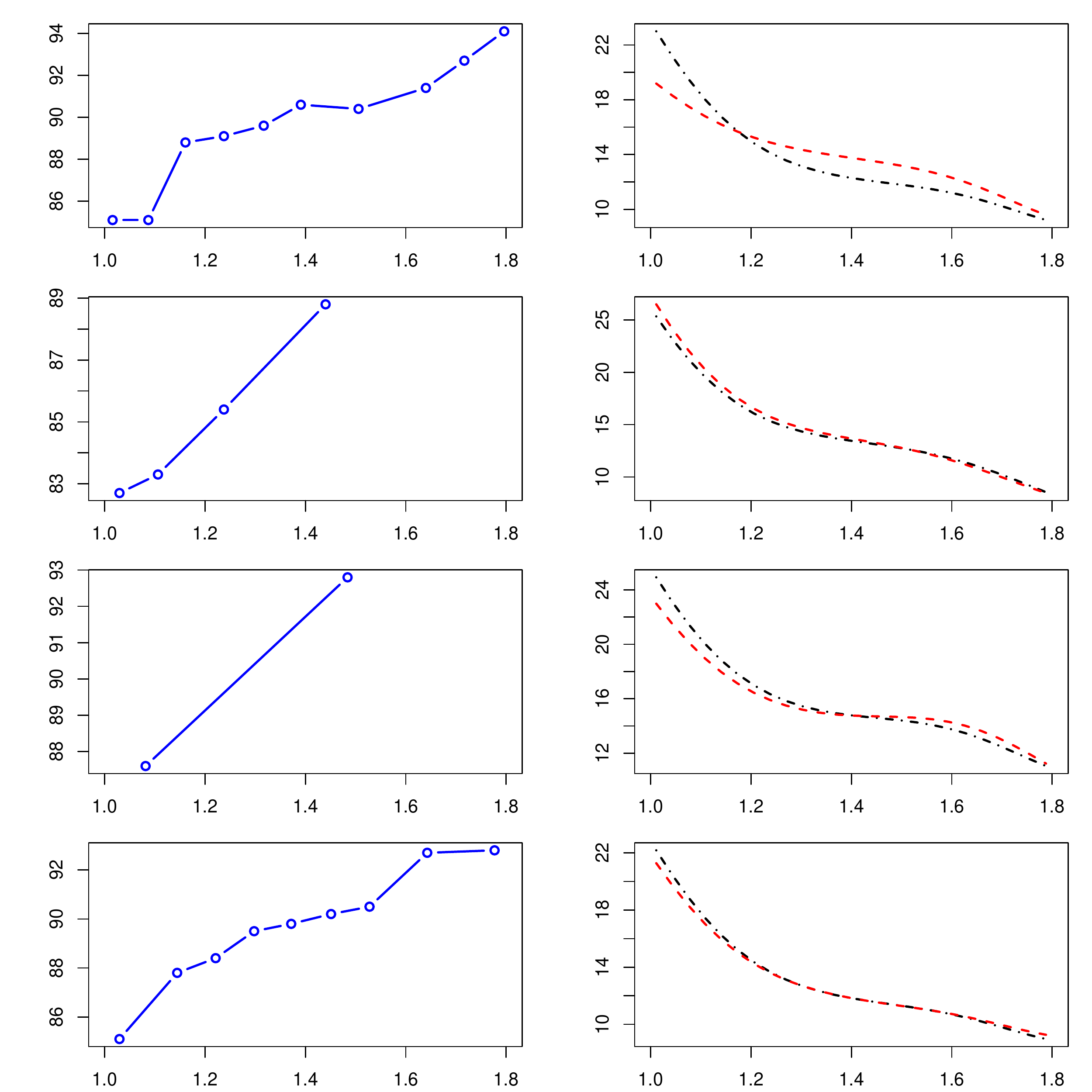}
\single \caption{\small{Data and corresponding estimated derivatives  of body length growth for  four randomly selected wallabies. Left panels: original body length data connected by lines; right panels: the derivative estimates obtained by FPCA  (dash-dot) and the proposed DPCA  (dashed). }}
\label{Wallaby_deriv}
\end{center}
\end{figure}

\autoref{Wallaby_deriv} exhibits the trajectories and corresponding derivatives estimates by FPCA and DPCA for four randomly selected wallabies, where the derivatives were constructed using $K=3$ components, the smallest number that leads to 90\% FVE. Here the DPC-derived derivatives are seen to be reflective of the data dynamics. % The confidence intervals constructed by DPC indicate that  the derivative estimates are more reliable in the initial period from 1 year to 1.4 years of age, while they are less stable for the later period from 1.4 years to 1.8 years of age. %The confidence intervals were narrower if more observations were available for the wallabies.

%\vspace{-0.5in}

\subsection{\sl Classifying Wheat Spectra} \label{ss:wheat}

\begin{figure}[h!]
\centering
\includegraphics[width=.9\textwidth]{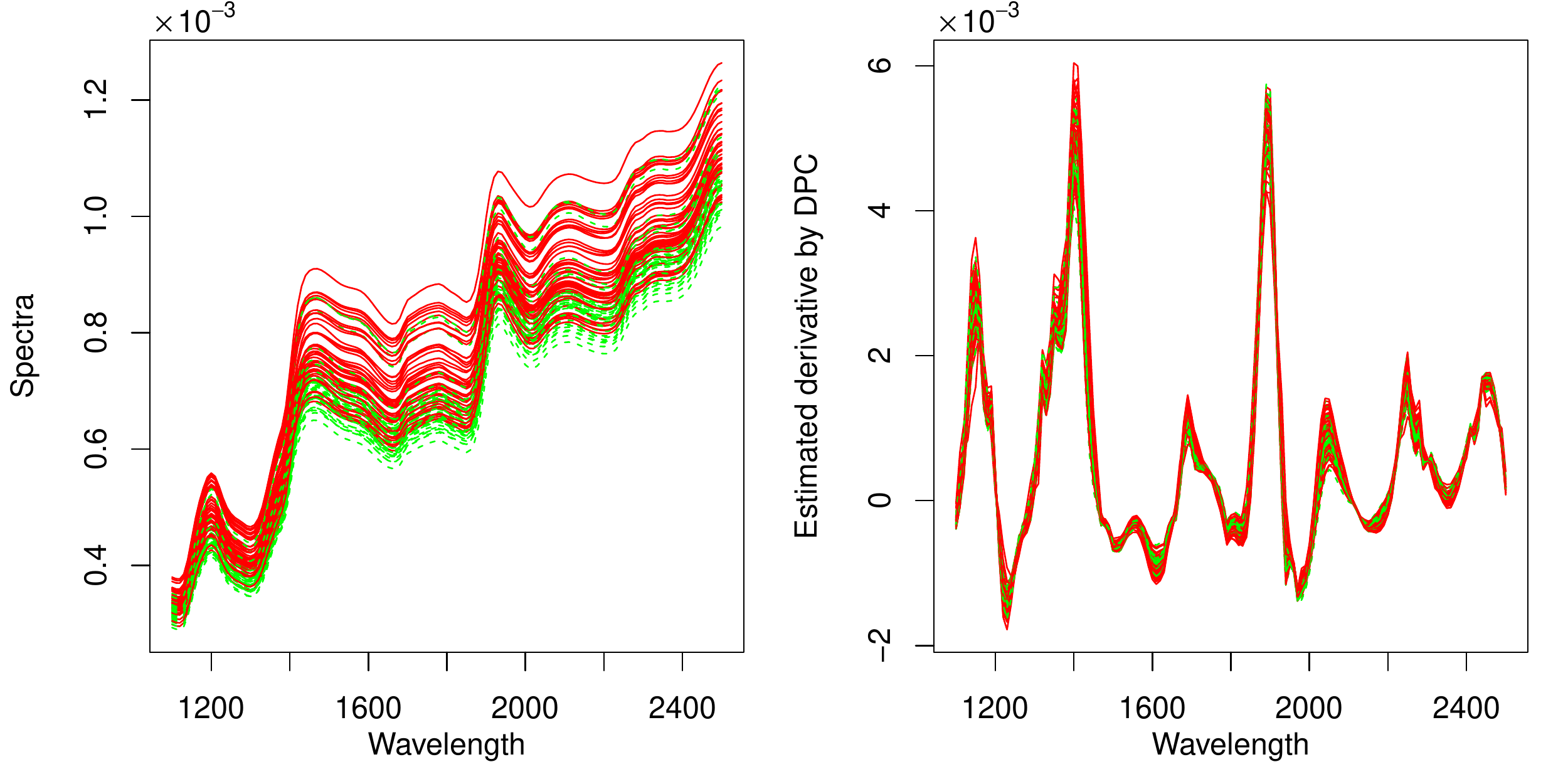}
\single \caption{\small{Left panel: observed trajectories for the NIR spectra of 87 wheat samples (high: solid; low: dashed). Right panel: estimated derivatives of the wheat sample NIR spectra trajectories using DPCA, based on the first four DPCs.}}
\label{wheat_trajectories}
\end{figure}

As a second example we applied the proposed  DPCA to the near infrared (NIR) spectral dataset of wheat, which consists of NIR spectra of 87 wheat samples with known protein content. The spectra of the samples were measured by diffuse reflectance from 1100 to 2500 nm with 10 nm intervals, as displayed in the left panel of \autoref{wheat_trajectories}. For these data, it is of interest to investigate whether the spectrum of a wheat sample can be utilized to  predict its protein content. Protein content is an important factor for wheat storage, and higher protein contents may increase the market price. For a more detailed description of these data, we refer to \ci{kali:97}.  Functional data analysis of these data has been studied by various authors, including \ci{reis:07}, and \ci{hall:12:2}.

As can be seen from the left panel of \autoref{wheat_trajectories}, the wheat samples are found to exhibit very similar spectral patterns: the overall trend for all trajectories is increasing, with three major local peaks appearing at wavelengths around 1200 nm, 1450 nm, and 1950 nm. The 
trajectories are almost parallel to each other, with only minor differences in the
overall level. The response to be predicted is the protein content of a wheat sample, which is grouped into categories--- high if a sample has more than 11.3\% protein, and low if less than 11.3\%. From the trajectory graphs it appears to be a non-trivial problem to classify the wheat samples, since the trajectories corresponding to the high protein wheats are highly spread out vertically and overlap on the lower side with those corresponding to the low group.

It has been suggested \citep{hall:12:2} that derivatives of wheat spectra are particularly suitable for classification of these spectra. We therefore applied  the proposed DPCA for the fitting of these spectra and this led to the estimated derivatives of wheat spectra shown in the right panel of \autoref{wheat_trajectories}. These fits are  based on including the first four DPCs, which
collectively explain 99.2\% of the total variation in the derivatives.

\vspace{.3in}
\begin{table}[ht]
\single
\centering
\caption{The mean fraction of misclassified samples, by randomly taking 30 samples as the training set and the rest 57 samples as the test set. The standard deviations of the misclassification rates are between 0.05 and 0.07. Classification models were built with different numbers of FPCs and DPCs, respectively. The first 8 columns are for fixed $K$ ranging from 1 to 8; the last column corresponds to selecting $K$ by 5-fold CV, with the mean of the chosen $K$ in brackets.} 
\label{table:wheat}
\begin{tabular}{|c|cccccccc|c|}
  \hline
 & $K=1$ & $K=2$ & $K=3$ & $K=4$ & $K=5$ & $K=6$ & $K=7$ & $K=8$ & CV \\ 
  \hline
FPCA & 0.274 & \textbf{0.253} & 0.267 & 0.284 & 0.293 & 0.280 & 0.273 & 0.280 & 0.282 (3.5) \\ 
  DPCA & 0.503 & \textbf{0.238} & 0.249 & 0.259 & 0.274 & 0.284 & 0.294 & 0.299 & 0.264 (3.5) \\ 
   \hline
\end{tabular}
\end{table}

For a comparative evaluation of the performance of using FPCA as opposed to DPCA
for the purpose of classifying protein contents of wheat samples, we used a logistic regression with one to eight FPCs or DPCs as predictors. We randomly drew 30 samples as training sets and 57 as test sets, repeated this 500 times, and report the average misclassification rates for the test sets in \autoref{table:wheat}, in which the first eight columns stand for using a fixed number of components, and the last column for selecting $K$ based on 5-fold cross-validation (CV), minimizing the misclassification rate. We found that the DPCA-based classifier outperforms the FPCA-based
classifier if two to five predictor components were included, or if CV was used to select $K$. The minimal misclassification rate is 23.8\% using two components, while the best FPCA-based misclassification rate is 25.3\%. The poor performance of DPCA when $K=1$ indicates the first DPC does not provide information for classification alone, while the second DPC may be a superior predictor of interest. While there are some improvements in the misclassification rates when using DPCs, they are relatively small in terms of misclassification error.

%\bc {\bf \sf  %ONLINE SUPPLEMENT
%\stepcounter{section}

\Appendix
\section{Estimating Mean Function and Eigenfunctions For Derivatives} \label{app:meanEF}

To implement \eqref{eq:XiK1},  one needs to  obtain $\mu^{(1)}(t)$, the first order derivative of the mean function. Let $N = \sumin N_i$, $w_i = N^{-1}$, and $v_i = [\sumin N_i (N_i - 1)]^{-1}$. Applying local quadratic smoothing to a pooled scatterplot, one can aggregate the observed measurements from all sample trajectories and minimize
\begin{equation}
\label{eq:muTarg}
\displaystyle \sum_{i=1}^{n} w_i \sumjNi \Khmu\Big(\frac{T_{ij}-t}{h_{\mu,1}}\Big)\Big[Y_{ij}- \sum_{p=0}^2 \alpha_p (T_{ij} - t)^p \Big]^2
\end{equation}
with respect to $\alpha_p$, $p=0,1,2$. The minimizer   $\hat{\alpha}_1(t)$ is the estimate of $\mu^{(1)}(t)$,  $\hat{\mu}^{(1)}(t)=\hat{\alpha}_1(t)$ (\cite{mull:09:1}, equation (5)).  Here $K_h(x) = h^{-1}K(x / h)$, $K(\cdot)$ is a univariate density function, and $h_\mu$ is a positive bandwidth that can be chosen by GCV in practical implementation.

In order to estimate the eigenfunctions $\phi_{k,1}$ of the derivatives $X^{(1)}$, we proceed by first estimating the covariance kernel $G_1(s,t)$ (in \eqref{eqn:sdG11} with $\nu=1$),  
followed by a spectral decomposition of the estimated kernel. According to \eqref{eqn: G11}, $\displaystyle G_{1}(s,t)=\Gone(s, t)$ for the case $\nu=1$; this can be estimated by a two-dimensional kernel smoother targeting the mixed partial derivatives of the covariance function. Specifically, we aim at minimizing 
\begin{equation}
\label{eq:GTarg}
\displaystyle \sumin v_i \sumjl \KhG(\Tijt) \KhG(\Tils) \Big[ G_i(T_{ij}, T_{il}) - \sum_{0 \le p + q \le 3} \alpha_{pq} (\Tijt)^p(\Tils)^q \Big]^2,
\end{equation}
with respect to $\alpha_{pq}$ for $0 \le p + q \le 3$, and set $\hGone(s,t)$ to the minimizer $\halpha_{11}(s, t)$. For theoretical derivations we adopt this direct estimate of $\hGone(s,t)$, while in practical implementation it has been found to be more convenient to first obtain $\hGonezero(s, t)$ and then to apply a local linear smoother on the second direction, which also led to better stability and boundary behavior. 
%Under certain  regulatory conditions,  $G^{(1,1)}(t,s)$ is equal to the second order partial derivative of  the covariance function $G(t,s)$, as
%\begin{align}
%G^{(1,1)}(t,s) &=E\{X_i^{(1)}(t)-\mu^{(1)}(t)\}\{X_i^{(1)}(s)-\mu^{(1)}(s)\} \nonumber \\
                     %   &= E\frac{\partial}{\partial t}\frac{\partial}{\partial s}\{X_i(t)-\mu(t)\} \{X_i(s)-\mu(s)\}  \nonumber \\
                      %  &=\frac{\partial}{\partial t}\frac{\partial}{\partial s}\text{E}\{X_i(t)-\mu(t)\}\{X_i(s)-\mu(s)\} \nonumber \\
                       % &=\frac{\partial^2}{\partial s \partial t}G(t,s)  \label{eqn: G11}
%\end{align}

%Using $G_{(1,1)}$ as kernel in a linear operator leads to the Hilbert-Schmidt operator, the ordered eigenfunctions and corresponding orthonormal eigenfunctions of which are denoted by $ \lambda_{1,1}> \lambda_{2,1}... \geq0 $ and $\phi_{k,1}(t)$, respectively. It is well known that $ G_{(1,1)}(s,t)$ has the spectral decomposition  as $ G_{(1,1)}(s,t)=  \sum_{k=1}^{\infty}\lambda_{k,1}\phi_{k,1}(s)\phi_{k,1}(t)$, and the derivative trajectories generated by this process has the Karhunen--Lo\`eve representation
%$X_i^{(1)}(t) = \mu^{(1)}(t) +\sum_{k=1}^{\infty}\xi_{ik,1}\phi_{k,1}(t)$. Here $\xi_{ik,1}=\int_{\mathcal{T}}(X^{(1)}_i(t)-\mu^{(1)}(t))\phi_{k,1}(t)dt$.

After obtaining estimates $\hat{G}^{(1, 1)}$ of $G^{(1,1)}$,  eigenfunctions and eigenvalues  of the derivatives  can be estimated by the spectral decomposition of $\hGone(s,t)$,
\[ \displaystyle \hat{G}^{(1, 1)}(s,t)= \sum_{k=1}^{\infty}\hat{\lambda}_{k,1}\hat{\phi}_{k,\nu}(s)\hat{\phi}_{k,1}(t).\]

\section{Assumptions, Proofs and Auxiliary  Results} \label{app:proofs}
For our results we require assumptions \autoref{a:K}--\autoref{a:Ubeta}, paralleling assumptions (A1)--(A2), (B1)--(B4), (C1c)--(C2c), and (D1c)--(D2c) in \cite{zhan:16}. Denote the inner product on $L^2(\cT)$ by $\inner{x}{y}$.

\begin{enumerate}[label=(A\arabic*)]
\item \label{a:K} $K(\cdot)$ is a symmetric probability density function on $[-1, 1]$ and  is Lipschitz continuous: There exists $0 < L < \infty$ such that $|K(u) - K(v)| \le L|u - v|$ for any $u, v \in [0, 1]$. %, which implies $K(\cdot)$ is bounded. 

\item \label{a:T} $\{ T_{ij}: i=1, \dots, n,\, j=1, \dots, N_i \}$ are i.i.d. copies of a random variable $T$ defined on $\cT$, and $N_i$ are regarded as fixed. The density $f(\cdot)$ of $T$ is bounded below and above,
\[ 0 < m_f \le \min_{\tinT} f(t) \le \max_{\tinT} f(t) \le M_f < \infty. \]
Furthermore $f^{(2)}$, the second derivative of $f(\cdot)$, is bounded.
\item \label{a:indep} $X$, $e$, and $T$ are independent.
\item \label{a:muGDerBounded} $\mu^{(3)}(t)$ and $\partial^4 G(s, t)/\partial^p\partial^{4-p}$ exist and are bounded on $\cT$ and $\cT \times \cT$, respectively, for $p = 0, \dots, 4$. 
\item \label{a:hmu} $h_\mu \tozero$ and $\log(n) \sumin N_i w_i^2/h_\mu \tozero$. 
\item \label{a:Ualpha} For some $\alpha > 2$, $E(\supt|X(t) - \mu(t)|^\alpha) < \infty$, $E(|e|^\alpha) < \infty$, and 
\[ n\left[ \sumin N_iw_i^2h_\mu + \sumin N_i(N_i - 1)w_i^2h_\mu^2 \right] \left[\frac{\log(n)}{n}\right]^{2/\alpha - 1} \toinf. \]
\item \label{a:hG} $h_G \tozero$, $\log(n)\sumin N_i(N_i - 1) v_i^2 / h_G^2 \tozero$. 
\item \label{a:Ubeta} For some $\beta > 2$, $E (\supt |X(t) - \mu(t)|^{2\beta}) < \infty$, $E(|e|^{2\beta}) < \infty$, and
\begin{align*}
n\bigg[ & \sumin N_i(N_i - 1)v_i^2h_G^2 + \sumin N_i(N_i-1)(N_i-2)v_i^2h_G^3 \\
	& + \sumin N_i(N_i-1)(N_i-2)(N_i-3)v_i^2h_G^4 \bigg]\left[ \frac{\log(n)}{n} \right]^{2/\beta - 1} \toinf.
\end{align*}
\item \label{a:phi1Finite} For any $k = 1, 2, \dots$, there exists $J = J(k) < \infty$ such that $\inner{\phi_j^{(1)}}{\phi_{k,1}}$ = 0 for all $j > J$. 
\end{enumerate}

%Let $\omega_K(s,t)=\bphi_{K,1,t}^T\Omega_K\bphi_{K,1,s}$ for $s, t \in \mathcal{T}$, $\bphi_{K,1,t}=(\phi_{1,1}(t),\phi_{2,1}(t),\ldots, \phi_{K,1}(t))^T$ and $\Omega_K=\text{var}(\tbxi_{iK,1}-\bxi_{iK,1})=\bLambda - \bH^T\SYiinv\bH$, where the definition of $\bH$ and $\bLambda$ are given before \autoref{thm:norm}. 
%Then $\{\omega_K(s,t)\}_{K=1}^\infty$ is a sequence of continuous positive definite functions. For the convergence of DPCs and the fitted derivative trajectories, we make the following assumptions, where \autoref{a:phi1Finite} pertains to the dense case only for \autoref{thm:xi}, and \autoref{a:omega} pertains to the sparse case only for \autoref{thm:norm}.
%\begin{enumerate}[label=(B\arabic*)]
%%\item \label{a:W} 
%\item \label{a:omega} There exists a continuous positive definite function $\omega(s,t)$ such that the sequence of functions $\omega_K$ satisfies $\omega_K(s,t)\rightarrow 	\omega(s,t) $ when $K \rightarrow \infty$.
%\end{enumerate}

Note that \autoref{a:phi1Finite} holds for any infinite-dimensional processes where  the eigenfunctions correspond to the Fourier basis or Legendre basis. 

\begin{proof}[Proof of \autoref{thm:muG}:]
We first prove the rate of convergence in the supremum norm for $\hmuone$ to $\muone$, following the proof of Theorem~5.1 in \cite{zhan:16}. Denote for $r = 0, \dots, 4$
\[
S_r = \sumin w_i \sumjNi \Khmu(\Tijt)\left(\Tijthmu\right)^r, \quad R_r = \sumin w_i \sumjNi \Khmu(\Tijt) \left(\Tijthmu \right)^r Y_{ij},
\]
\[
\bS = \left[ \begin{array}{ccc}
S_0 & S_1 & S_2 \\ 
S_1 & S_2 & S_3 \\ 
S_2 & S_3 & S_4
\end{array}  \right], \quad 
\left[
\begin{array}{c}
\halpha_0 \\ h_\mu\halpha_1 \\ h_\mu^2\halpha_2
\end{array}
\right] = \bS^{-1} 
\left[\begin{array}{c}
R_0 \\ R_1 \\ R_2
\end{array}\right].
\]
For a square matrix $\bA$ let $|\bA|$ denote its determinant and $[\bA]_{a,b}$ denote the $(a,b)$th entry of $\bA$. Then $h_\mu\hmuone(t) = h_\mu\halpha_0 = |\bS|^{-1}(C_{12}R_0 + C_{22}R_1 + C_{32}R_2)$ by Cramer's rule \citep{lang:87}, where 
\[
C_{12} = \left| \begin{array}{cc}
S_1 & S_3 \\
S_2 & S_4
\end{array}\right|, \quad 
C_{22} = \left|\begin{array}{cc}
S_0 & S_2 \\
S_2 & S_4
\end{array}\right|, \quad
C_{23} = \left|\begin{array}{cc}
S_0 & S_2 \\
S_1 & S_3
\end{array}\right|
\]
are the cofactors for $[\bS]_{1, 2}$, $[\bS]_{2,2}$, and $[\bS]_{3, 2}$, respectively. Then 
\begingroup \allowdisplaybreaks
\begin{align*}
h_\mu(\halpha_1 & - \muone(t)) = |\bS|^{-1}\{\left[C_{12}R_0 + C_{22}R_1 + C_{32}R_2 \right] \\
& \hspace{6em} -\left[C_{12}S_0 + C_{22}S_1 + C_{32}S_2 \right] \\
& \hspace{6em} - \left[C_{12}S_2 + C_{22}S_3 + C_{32}S_4 \right]\mutwo(t)h_\mu^2\\
& \hspace{6em} - \left[C_{12}S_1 + C_{22}S_2 + C_{32}S_3 \right]\muone(t)h_\mu\} \\
& = |\bS|^{-1} \sum_{p=0}^2 C_{(p+1), 2}(R_p - S_p - \muone(t)h_\mu S_{p+1} - \mutwo(t) h_\mu S_{p+2}) \\
& = |\bS|^{-1} \sum_{p=0}^2 C_{(p+1), 2}\bigg[ \sumin w_i \sumjNi \Khmu(\Tijt)\left(\Tijthmu\right)^p\delta_{ij} \\
& \quad + \sumin w_i \sumjNi \Khmu(\Tijt) \left(\Tijthmu\right)^p h_\mu^3\muthree(z) \bigg] \\
& = |\bS|^{-1}\sum_{p=0}^2 C_{(p+1),2} \bigg[O\bigg(\bigg\{\log(n)\bigg[\sumin N_iw_i^2h_\mu + \sumin N_i(N_i-1)w_i^2h_\mu^2 \bigg] \bigg\}^{1/2}\bigg) + O(h_\mu^3)\bigg] \quad \almostsure;
\end{align*}\endgroup
here the first equality is due to the properties of determinants, the third is due to Taylor's theorem, and the last is due to Lemma~5 in \cite{zhan:16}, \autoref{a:K}, and \autoref{a:muGDerBounded}, where the  $O(\cdot)$ terms are seen to be  uniform in $\tinT$. By Theorem~5.1 in \cite{zhan:16}, the $S_r$ converge almost surely to their respective means in supremum norm and are thus bounded almost surely for $r = 0, \dots, 4$, so that $C_{(p+1), 2}$ is bounded almost surely for $p = 0, 1, 2$. Then $|\bS|^{-1}$ is bounded away from 0 by the almost sure supremum convergence of $S_r$ and Slutsky's theorem. Therefore the convergence rate for $\hmuone$ is
\[
\supt |\hmuone(t) - \muone(t)| = O\bigg(\bigg\{\log(n)\bigg[\sumin N_iw_i^2/h_\mu + \sumin N_i(N_i-1)w_i^2 \bigg] \bigg\}^{1/2} + h_\mu^2\bigg) \quadas
\]
The rate \eqref{eq:mu1Rt} then follows by replacing $N_i$ by $N_0$ in the sparse case where $N_i \le N_0 < \infty$, and by $m$ in the dense case where $m \toinf$, respectively. 

The supremum convergence rate for $\hGone$ can  be proven similarly, following the development of Theorem~5.2 in \cite{zhan:16}. The supremum convergence rate for $\hphikone$ is a direct consequence of that for $\hGone$; see the proof of Theorem~2 in \cite{mull:05:4}. 
\end{proof}

\begin{proof}[Proof of \autoref{thm:xi}:]
For a vector $\bv$, let $\norm{\bv}$ be the vector $L^2$ norm and, for a square matrix $\bA$, let $\norm{\bA} = \sup_{\bv\ne 0} \norm{\bA\bv}/\norm{\bv}$ be the matrix operator norm. We start with  a proposition that follows from the proof of Corollary~1 in \cite{mull:05:4},  our \autoref{thm:muG}, and a lemma from \cite{mull:03:9}.
\begin{prop} \label{prop:muG}
Under the conditions of \autoref{thm:muG}, 
\begin{align*}
\supt|\hmu(t) - \mu(t)| & = O(a_n) \quadas,  \\
\supst|\hG(s, t) - G(s, t)| & = O(a_n + b_n) \quadas, \\
\supst|\hGonezero(s, t) - \Gonezero(s, t)| & = O(a_n + b_n) \quadas, \\
|\hsigma^2 - \sigma^2| & = O(a_n + b_n) \quadas
\end{align*}
\end{prop}

\begin{lem}[Lemma A.3, \cite{mull:03:9}] \label{lem:invDiff} 
Let $\bA \in \mathcal{M}_m(\bbR) $ be invertible. For all $\bB \in \mathcal{M}_m(\bbR)$ such that \[\|\bA-\bB\|<\frac{1}{2\|\bA^{-1}\|}, \] $\bB^{-1}$ always exists and there exists a constant $0<c<\infty$ such that
\[\|\bB^{-1}-\bA^{-1}\|\leq c\|\bA^{-1}\|^2\|\bA-\bB\|.\]
\end{lem}
	
To prove the first statement of \autoref{thm:xi}, note
\begin{align}
|\hxiikone - \txiikone| & = \hbzetaik^T\hSYiinv(\bYi - \hbmui) - \bzetaik^T \SYiinv (\bYi - \bmui) \nonumber \\
& = \hbzetaik^T(\hSYiinv - \SYiinv)(\bYi - \hbmui) + (\hbzetaik - \bzetaik)^T\SYiinv(\bYi-\hbmui) \nonumber\\
& \quad + \hbzetaik\SYiinv(\bmui - \hbmui) - (\hbzetaik - \bzetaik)^T\SYiinv(\bmui - \hbmui) \nonumber\\
& \le \norm{\hbzetaik}\norm{\hSYiinv - \SYiinv} \norm{\bYi - \hbmui} + \norm{\hbzetaik - \bzetaik}\norm{\SYiinv}\norm{\bYi-\hbmui} \nonumber\\
& \quad + \norm{\hbzetaik}\norm{\SYiinv}\norm{\bmui - \hbmui} + \norm{\hbzetaik - \bzetaik}\norm{\SYiinv}\norm{\bmui - \hbmui}. \label{eq:xihtDiff}
\end{align}

We bound each term as follows, using the notation  $\lesssim$ to indicate that the left hand side is smaller than a constant multiple of the right hand side. We have $\norm{\hbzetaik - \bzetaik} \le \sqrtNi\supj |\hzetaikj - \zetaikj|$, and 
\begin{align*}
\supj |\hzetaikj - \zetaikj| & = \supj |\int \hGonezero(s, T_{ij})\hphikone(s) ds - \int \Gonezero(s, T_{ij})\phikone(s) ds| \\
& \le \supt |\int [\hGonezero(s, t) - \Gonezero(s, t)]\hphikone(s) ds| + \supt \int \Gonezero(s, t)[\hphikone(s) - \phikone(s)] ds \\
& \lesssim \supt \left\{\int [\hGonezero(s, t) - \Gonezero(s, t)]^2 ds \right\}^{1/2} + \supst |\Gonezero(s, t)| \supt|\hphikone(t) - \phikone(t)| \\
& = O(\supt|\hGonezero-\Gonezero|) + O(\supt|\hphikone(t) - \phikone(t)|),
\end{align*}
where the last equality is due to \autoref{a:muGDerBounded}. By \autoref{prop:muG} we have
\begin{equation} \label{eq:zetaDiffRt}
\norm{\hbzetaik - \bzetaik} = O(\sqrtNi (a_n+b_n)) \quadas
\end{equation}
Similarly, $\supj |\zetaikj| \le \supt|\Gonezero(s,t)\phikone(s) ds| = O(1)$. Take $\beps_i = (\eps_{i1}, \dots, \eps_{im})^T$ and $\bXi = (X(T_{i1}), \dots, X(T_{im}))^T$. Then 
\begin{align}
\norm{\hbzetaik} & \le \norm{\bzetaik} + \norm{\hbzetaik - \bzetaik} = \sqrtNi [O(1) + O(a_n + b_n)] =  O(\sqrtNi) \quadas, \label{eq:zetaRt}\\
\norm{\bmui - \hbmui} & \le \sqrtNi\supt|\hmu(t) - \mu(t)| = O(\sqrtNi a_n) \quadas, \label{eq:bmuRt}\\
\norm{\bYi - \hbmui} & \le \norm{\beps_i}  + \norm{\bXi - \bmui} + \norm{\hbmui - \bmui}  =  O_p(\sqrtNi) \label{eq:YiRt}
\end{align}
where \eqref{eq:YiRt} is by the Weak Law of Large Numbers. From the definition of $\SYi$ we have $\norm{\SYiinv} \le \sigma^{-2}$. Then 
\begin{align}
\norm{\hSYiinv - \SYiinv} & \le c\sigma^{-4}\norm{\hSYi - \SYi} \nonumber \\
& \le c\sigma^{-4} N_i \supab|\hSYiab - \SYiab| \nonumber\\
& = O(N_i(a_n + b_n)) \quadas, \label{eq:SYiRt}
\end{align}
where the first inequality is by \autoref{lem:invDiff}, the second by a property of matrix operator norm, and the last relies on $\supab|\hSYiab - \SYiab| \le |\hsigma^2 - \sigma^2| + \supst|\hG(s,t) - G(s,t)| = O(a_n + b_n)$ a.s. by \autoref{prop:muG}.
Combining \eqref{eq:xihtDiff}--\eqref{eq:SYiRt} leads to the proof of the first statement. 

Under the dense assumption we have $N_i = m \toinf$. Since $\zeta_{ikl} = \int_\cT \Gonezero(s, T_{il}) \phikone(s) ds = \int_{\cT} \sumjinf \lambda_j \phi_j^{(1)}(s)\phi_j(T_{il}) \phikone(s) ds = \sumjinf \lambda_j\inner{\phi_j^{(1)}}{\phikone}\phi_j(T_{il})$ for $l = 1, \dots, m$, under \autoref{a:phi1Finite} we have $\bzetaik = \sumjJ\lambda_j\inner{\phi_j^{(1)}}{\phikone}\bphi_j$, where we take $\bphi_j = (\phi_j(T_{i1}), \dots, \phi_j(T_{im}))^T$. Then
\begin{align*}
\txiikone & = \bzetaik^T\SYiinv(\bYi - \bmui) = \sumjJ\lambda_j\inner{\phi_j^{(1)}}{\phikone}\bphi_j^T\SYiinv(\bYi - \bmui), \\
\xiikone & = \inner{\Xone}{\phikone} = \inner{\sumjinf\xi_{ij}\phi_j^{(1)}}{\phikone} = \sumjJ\xi_{ij}\inner{\phi_j^{(1)}}{\phikone},
\end{align*}
so it suffices to prove for $j = 1, \dots, J$, 
\begin{equation} \label{eq:phijInner}
\bphi_j^T\SYiinv(\bYi - \bmui) = \xi_{ij}/\lambda_j + O_p(m^{-1/2}).
\end{equation}
Under joint Gaussianity of $(X, \epsilon)$, $E(\xi_{ij} \mid \bYi) = \lambda_j \bphi_j^T\SYiinv(\bYi - \bmui)$ is the posterior mean of $\inner{X_i}{\phi_j}$ given the observations $\bYi$. By the convergence results for nonparametric posterior distributions as in Theorem~3 of \cite{shen:02}, we have 
\[
|E(\xi_{ij} \mid \bYi) -  \xi_{ij}| = O_p(m^{-1/2})
\]
as $m \toinf$, which implies \eqref{eq:phijInner} and therefore the second statement of \autoref{thm:xi}. 

\end{proof}

\begin{proof}[Proof of \autoref{thm:X}:]
For all $i=1,2,\ldots$ and any fixed $K$,
\begin{align}\label{Xik_est_condition}
\normSup{\hat{X}^{(1)}_{i,K}(t)- \tilde{X}^{(1)}_{i,K}(t)} 
&=\normSup{\sum_{k=1}^K(\hat{\xi}_{ik,1}-\tilde{\xi}_{ik,1})\hat{\phi}_{k,1}(t)+\sum_{k=1}^K\tilde{\xi}_{ik,1}(\hat{\phi}_{k,1}(t)-\phi_{k,1}(t))}\nonumber\\
&= \sum_{k=1}^K|\hat{\xi}_{ik,1}-\tilde{\xi}_{ik,1}| O_p(1)+\sum_{k=1}^K|\tilde{\xi}_{ik,1}| \, \normSup{\hat{\phi}_{k,1}(t)-\phi_{k,1}(t)} \nonumber\\
& = O_p(N_i^2(a_n + b_n) + (a_n + b_n)) = O_p(N_i^2(a_n + b_n)). \nonumber
\end{align}

A similar rate for $\normSup{\hXone_{i,K}(t) - \Xone_{i,K}(t)}$ in the dense case is obtained by applying \autoref{thm:xi} and repeating the previous argument. 
\end{proof}

%\subsubsection*{\it Proof of \autoref{thm:norm}:}
%Similar to the proof of Theorem 3 in \cite{mull:05:4}.

\references

\end{document}

%% file: preamble.tex
\usepackage{natbib,graphics,geometry,epsfig,rotate,lscape,
graphicx,amsmath,colortbl, xspace, float,
amsthm,amssymb,sectsty,amsfonts,amsbsy,bm,color,multirow,mathtools}
\usepackage{enumitem}
\usepackage[colorlinks=true, allcolors=blue]{hyperref}

\def\references{\bibliography{deriv}}

\bibpunct{(}{)}{;}{a}{}{,}

\renewcommand{\citep}[1]{(\cite{#1})}

\sectionfont{\centering\bf\sf\normalsize}
\subsectionfont{\sf\normalsize}

\newcommand{\Appendix}{\appendix
\section*{Appendix}
\def\thesection{A.\arabic{section}}
\def\thesubsection{\Alph{section}.\arabic{subsection}}}

\renewcommand{\baselinestretch}{1.6} %{2} %{1.4}
\newcommand{\single}{\renewcommand{\baselinestretch}{1}\normalsize}
\newcommand{\double}{\renewcommand{\baselinestretch}{2}\normalsize}

  \renewenvironment{thebibliography}[1]{
    \begin{oldthebibliography}{#1}
      \setlength{\parskip}{0ex}
      \setlength{\itemsep}{0ex}  }{    \end{oldthebibliography} }

\newcommand{\bea}{\begin{eqnarray*}}
\newcommand{\eea}{\end{eqnarray*}}
\newcommand{\ed}{\end{document}}

\newcommand{\btab}{\begin{tabular}}
\newcommand{\etab}{\end{tabular}}

\newcommand{\bi}{\begin{itemize}}
\newcommand{\ei}{\end{itemize}}
\newcommand{\bfi}{\begin{figure}}
\newcommand{\efi}{\end{figure}}
\newcommand{\ben}{\begin{enumerate}}
\newcommand{\een}{\end{enumerate}}
\newcommand{\bay}{\begin{array}}
\newcommand{\eay}{\end{array}}

\definecolor{DarkBlue}{rgb}{0,.08,.45}
\definecolor{DarkRed}{rgb}{.7,0,.4}

\def\bco{\iffalse}
\def\cov{{\rm cov}}
\def\var{{\rm var}}

\def\ci{\cite}
\newcommand{\cp}{\citep}
\def\eps{\varepsilon}

\newtheorem{thm}{Theorem}
\newtheorem{lem}{Lemma}

\newtheorem{prop}{Proposition}

\newcommand{\no}{\noindent}
\newcommand{\bc}{\begin{center}}
\newcommand{\ec}{\end{center}}

\newcommand*{\bbR}{\mathbb{R}}
\newcommand*{\cT}{\mathcal{T}}

\newcommand{\Xone}{X^{(1)}}
\newcommand{\hXone}{\hX^{(1)}}
\newcommand{\sumin}{\sum_{i=1}^n}

\newcommand{\muone}{\mu^{(1)}}
\newcommand{\mutwo}{\mu^{(2)}}
\newcommand{\muthree}{\mu^{(3)}}

\newcommand{\phikone}{\phi_{k,1}}
\newcommand{\hphikone}{\hphi_{k,1}}

\newcommand{\hphi}{\hat{\phi}}
\newcommand{\hG}{\hat{G}}

\newcommand{\hX}{\hat{X}}
\newcommand{\hmu}{\hat{\mu}}
\newcommand{\hxi}{\hat{\xi}}
\newcommand{\hsigma}{\hat{\sigma}}
\newcommand{\hzeta}{\hat{\zeta}}
\newcommand{\halpha}{\hat{\alpha}}

\newcommand{\hlambda}{\hat{\lambda}}

\newcommand{\tX}{\tilde{X}}

\newcommand{\txi}{\tilde{\xi}}

\newcommand{\bX}{\mathbf{X}}
\newcommand{\bY}{\mathbf{Y}}

\newcommand{\bS}{\mathbf{S}}
\newcommand{\bA}{\mathbf{A}}
\newcommand{\bB}{\mathbf{B}}

\newcommand{\bmu}{\boldsymbol{\mu}}

\newcommand{\bzetaik}{\boldsymbol{\zeta}_{ik}}
\newcommand{\bzetaiknu}{\boldsymbol{\zeta}_{ik,\nu}}
\newcommand{\hbzetaik}{\hat{\boldsymbol{\zeta}}_{ik}}
\newcommand{\bphi}{\boldsymbol{\phi}}

\newcommand{\beps}{\boldsymbol{\epsilon}}

\newcommand{\bSigma}{\boldsymbol{\Sigma}}
\newcommand{\bmui}{\bmu_i}
\newcommand{\hbmui}{\hat{\bmu}_i}
\newcommand{\bYi}{\bY_i}
\newcommand{\bXi}{\bX_i}

\newcommand{\bv}{\mathbf{v}}

\newcommand{\toinf}{\rightarrow \infty}
\newcommand{\tozero}{\rightarrow 0}

\newcommand{\sumjinf}{\sum_{j=1}^\infty}
\newcommand{\sumjJ}{\sum_{j=1}^J}

\newcommand{\Gone}{G^{(1, 1)}}
\newcommand{\Gonezero}{G^{(1, 0)}}
\newcommand{\hGone}{\hG^{(1, 1)}}
\newcommand{\hGonezero}{\hG^{(1, 0)}}

\newcommand{\hmuone}{\hmu^{(1)}}
\newcommand{\hxiikone}{\hxi_{ik,1}}
\newcommand{\txiikone}{\txi_{ik,1}}
\newcommand{\txiiknu}{\txi_{ik,\nu}}
\newcommand{\xiikone}{\xi_{ik,1}}
\newcommand{\xiiknu}{\xi_{ik,\nu}}

\DeclarePairedDelimiterX{\inner}[2]{\langle}{\rangle}{#1, #2}

\newcommand{\Khmu}{K_{h_\mu}}
\newcommand{\KhG}{K_{h_G}}
\newcommand{\sumjl}{\sum_{1 \le j \ne l \le N_i}}
\newcommand{\sumjNi}{\sum_{j=1}^{N_i}}
\newcommand{\Tijt}{T_{ij}-t}
\newcommand{\Tils}{T_{il}-s}
\newcommand{\tinT}{t\in \cT}
\newcommand{\supt}{\sup_{t\in\cT}}
\newcommand{\supst}{\sup_{s,t\in\cT}}
\newcommand{\almostsure}{\text{a.s.}}
\newcommand{\quadas}{\quad \almostsure}
\newcommand{\SYi}{\bSigma_{\bYi}}
\newcommand{\hSYi}{\hat{\bSigma}_{\bYi}}
\newcommand{\SYiinv}{\bSigma_{\bYi}^{-1}}
\newcommand{\hSYiinv}{\hat\bSigma_{\bYi}^{-1}}
\newcommand{\hSYiab}{[\hSYi]_{ab}}
\newcommand{\SYiab}{[\SYi]_{ab}}

\newcommand{\supj}{\sup_{j}}
\newcommand{\supab}{\sup_{a,b}}

\newcommand{\Tijthmu}{\frac{\Tijt}{h_\mu}}

\newcommand{\norm}[1]{\left\lVert#1\right\rVert}

\newcommand{\normSup}[1]{\supt|#1|}
\newcommand{\sqrtNi}{\sqrt{N_i}}

\newcommand{\zetaikj}{\zeta_{ikj}}
\newcommand{\hzetaikj}{\hzeta_{ikj}}

%% file: stat-sinica2.bbl
\begin{thebibliography}{27}
\expandafter\ifx\csname natexlab\endcsname\relax\def\natexlab#1{#1}\fi
\expandafter\ifx\csname url\endcsname\relax
  \def\url#1{\texttt{#1}}\fi
\expandafter\ifx\csname urlprefix\endcsname\relax\def\urlprefix{URL }\fi

\bibitem[{Bapna \textit{et~al.}(2008)Bapna, Jank and Shmueli}]{bapn:08}
Bapna, R., Jank, W. and Shmueli, G. (2008). Price formation and its dynamics in
  online auctions.
\newblock \textit{Decision Support Systems} \textbf{44}, 641--656.

\bibitem[{de~Boor(1972)}]{debo:72}
de~Boor, C. (1972). On calculating with {$B$}-splines.
\newblock \textit{Journal of Approximation Theory} \textbf{6}, 50--62.

\bibitem[{Chambers and Hastie(1991)}]{cham:91}
Chambers, J.~M. and Hastie, T. (eds.) (1991). \textit{Statistical Models in
  {S}}.
\newblock Pacific Grove: Duxbury Press.

\bibitem[{Delaigle and Hall(2012)}]{hall:12:2}
Delaigle, A. and Hall, P. (2012). Achieving near perfect classification for
  functional data.
\newblock \textit{Journal of the Royal Statistical Society: Series B
  (Statistical Methodology)} \textbf{74}, 267--286.

\bibitem[{Facer and M\"{u}ller(2003)}]{mull:03:9}
Facer, M. and M\"{u}ller, H.-G. (2003). Nonparametric estimation of the peak
  location in a response surface.
\newblock \textit{Journal of Multivariate Analysis} \textbf{87}, 191--217.

\bibitem[{Fan and Gijbels(1996)}]{fan:96}
Fan, J. and Gijbels, I. (1996). \textit{Local Polynomial Modelling and its
  Applications}.
\newblock London: Chapman \& Hall.

\bibitem[{Gasser and M\"{u}ller(1984)}]{mull:84:1}
Gasser, T. and M\"{u}ller, H.-G. (1984). Estimating regression functions and
  their derivatives by the kernel method.
\newblock \textit{Scandinavian Journal of Statistics} \textbf{11}, 171--185.

\bibitem[{Grenander(1950)}]{gren:50}
Grenander, U. (1950). Stochastic processes and statistical inference.
\newblock \textit{Arkiv f\"or Matematik} \textbf{1}, 195--277.

\bibitem[{Jank and Shmueli(2005)}]{jank:05:1}
Jank, W. and Shmueli, G. (2005). Profiling price dynamics in online auctions
  using curve clustering.
\newblock \textit{Technical Report. SSRN eLibrary}.

\bibitem[{Kalivas(1997)}]{kali:97}
Kalivas, J.~H. (1997). Two data sets of near infrared spectra.
\newblock \textit{Chemometrics and Intelligent Laboratory Systems} \textbf{37},
  255--259.

\bibitem[{Lang(1987)}]{lang:87}
Lang, S. (1987). \textit{Linear Algebra}.
\newblock New York: Springer.

\bibitem[{Liu and M\"{u}ller(2009)}]{mull:09:1}
Liu, B. and M\"{u}ller, H.-G. (2009). Estimating derivatives for samples of
  sparsely observed functions, with application to on-line auction dynamics.
\newblock \textit{Journal of the American Statistical Association}
  \textbf{104}, 704--714.

\bibitem[{Mallon(1994)}]{mall:94}
Mallon, G. (1994). \textit{Mixed linear models and applications}.
\newblock Ph.D. thesis. University of Queensland, Department of Mathematics.

\bibitem[{M\"{u}ller and Yao(2010)}]{mull:10:2}
M\"{u}ller, H.-G. and Yao, F. (2010). Empirical dynamics for longitudinal data.
\newblock \textit{Annals of Statistics} \textbf{38}, 3458--3486.

\bibitem[{Ramsay and Silverman(2005)}]{rams:05}
Ramsay, J.~O. and Silverman, B.~W. (2005). \textit{Functional {D}ata
  {A}nalysis}.
\newblock New York: Springer. second edn.

\bibitem[{Reddy and Dass(2006)}]{redd:06}
Reddy, S.~K. and Dass, M. (2006). Modeling on-line art auction dynamics using
  functional data analysis.
\newblock \textit{Statistical Science} \textbf{21}, 179--193.

\bibitem[{Reiss and Ogden(2007)}]{reis:07}
Reiss, P. and Ogden, R. (2007). Functional principal component regression and
  functional partial least square.
\newblock \textit{Journal of the American Statistical Association}
  \textbf{102}, 984--996.

\bibitem[{Rice and Silverman(1991)}]{rice:91}
Rice, J.~A. and Silverman, B.~W. (1991). Estimating the mean and covariance
  structure nonparametrically when the data are curves.
\newblock \textit{Journal of the Royal Statistical Society: Series B}
  \textbf{53}, 233--243.

\bibitem[{Rice and Wu(2001)}]{rice:01}
Rice, J.~A. and Wu, C.~O. (2001). Nonparametric mixed effects models for
  unequally sampled noisy curves.
\newblock \textit{Biometrics} \textbf{57}, 253--259.

\bibitem[{Schwarz(1978)}]{schw:78}
Schwarz, G. (1978). Estimating the dimension of a model.
\newblock \textit{Annals of Statistics} \textbf{6}, 461--464.

\bibitem[{Shen(2002)}]{shen:02}
Shen, X. (2002). Asymptotic normality of semiparametric and nonparametric
  posterior distributions.
\newblock \textit{Journal of the American Statistical Association} \textbf{97},
  222--235.

\bibitem[{Shi \textit{et~al.}(1996)Shi, Weiss and Taylor}]{shi:96}
Shi, M., Weiss, R.~E. and Taylor, J. M.~G. (1996). An analysis of paediatric
  {CD}4 counts for {A}cquired {I}mmune {D}eficiency {S}yndrome using flexible
  random curves.
\newblock \textit{Journal of the Royal Statistical Society: Series C (Applied
  Statistics)} \textbf{45}, 151--163.

\bibitem[{Shibata(1981)}]{shib:81}
Shibata, R. (1981). An optimal selection of regression variables.
\newblock \textit{Biometrika} \textbf{68}, 45--54.

\bibitem[{Wang \textit{et~al.}(2008)Wang, Jank, Shmueli and Smith}]{wang:08:2}
Wang, S., Jank, W., Shmueli, G. and Smith, P. (2008). Modeling price dynamics
  in ebay auctions using principal differential analysis.
\newblock \textit{Journal of the American Statistical Association}
  \textbf{103}, 1100--1118.

\bibitem[{Yao \textit{et~al.}(2005)Yao, M\"{u}ller and Wang}]{mull:05:4}
Yao, F., M\"{u}ller, H.-G. and Wang, J.-L. (2005). Functional data analysis for
  sparse longitudinal data.
\newblock \textit{Journal of the American Statistical Association}
  \textbf{100}, 577--590.

\bibitem[{Zhang and Wang(2016)}]{zhan:16}
Zhang, X. and Wang, J.-L. (2016). From sparse to dense functional data and
  beyond.
\newblock \textit{The Annals of Statistics} \textbf{44}, 2281--2321.

\bibitem[{Zhou and Wolfe(2000)}]{zhou:00}
Zhou, S. and Wolfe, D.~A. (2000). On derivative estimation in spline
  regression.
\newblock \textit{Statistica Sinica} \textbf{10}, 93--108.

\end{thebibliography}
